\documentclass[aps,pra,twocolumn,groupedaddress,longbibliography]{revtex4-1}

\usepackage[normalem]{ulem}

\usepackage[colorlinks,bookmarks=false,citecolor=blue,linkcolor=red,urlcolor=blue]{hyperref}
\usepackage{enumitem}
\usepackage{graphicx}
\usepackage{amsmath}

\begin{document}

\title{Thermalization and its Breakdown for a Large Nonlinear Spin}

\author{Shane P. Kelly}
\email[]{skell013@ucr.edu}
\affiliation{Theoretical Division, Los Alamos National Laboratory, Los Alamos, New Mexico 87545, USA}
\affiliation{Department of Physics and Astronomy, University of California Riverside, Riverside, California 92521, USA}

\author{Eddy Timmermans}
\affiliation{XCP-5, XCP Division, Los Alamos National Laboratory, Los Alamos, New Mexico 87545, USA}

\author{S.-W. Tsai}
\affiliation{Department of Physics and Astronomy, University of California Riverside, Riverside, California 92521, USA}


\date{\today}

\begin{abstract}
    By developing a semi-classical analysis based on the Eigenstate Thermalization Hypothesis, we determine the long time behavior of a large spin evolving with a nonlinear Hamiltonian.
    Despite integrable classical dynamics, we find the Eigenstate Thermalization Hypothesis { for the diagonal matrix elements of observables} is satisfied in the majority of eigenstates, and thermalization of long time averaged observables is generic. 
    The exception is a novel mechanism for the breakdown of thermalization based on an unstable fixed point in the classical dynamics.
    Using the semi-classical analysis we derive how the equilibrium values of observables encode properties of the initial state.
    { This analysis shows an unusual memory effect in which the remembered initial state property is not conserved in the integrable classical dynamics.}
    We conclude with a discussion of relevant experiments and the potential generality of { this mechanism for long time memory and the breakdown of thermalization.}
\end{abstract}

\maketitle

%Clean up:
In recent years, experiments on ultra cold atoms and trapped ions\cite{bloch2008,ludlow2015,shaffer2018,zibold2010} have succeeded in producing quantum systems that,  on relevant time scales, are completely isolated from an environment.
Surprisingly, many of these experiments find long time behavior that mimics a system coupled to an environment.
These experiments prompt the question of thermalization:
Given an initial state $\left|\psi(t=0)\right>$, a Hamiltonian $H=\sum_{n}E_{n}\left| n\right>\left< n\right|$, and an observable $O$, when and why does the long time average of $O$:
\begin{eqnarray}
    O(t,T)=\frac{1}{T}\int_{t}^{t+T}d\tau \left<\psi(\tau)\right|O\left|\psi(\tau)\right>
\end{eqnarray}
lose memory of its initial state? In other words, when does $O(t,T)$, at long time $t$, depend only on the energy of the initial state?

The eigenstate thermalization hypothesis (ETH)\cite{deutsch1991,srednicki1994,rigol2008,dalessio2016a,deutsch2018,jensen1985, Mori_2018} attempts to answer this question.
Briefly, it states that if 
%\begin{enumerate}[label=A\arabic*)]
    A1) $\left<n\right|O\left|n\right>$ changes very little between eigenstates with similar energy;
    A2) the level spacings, $E_n-E_{n+1}$, are sufficiently small; 
    and A3) the energy uncertainty of the initial state is sufficiently small,
%\end{enumerate}
%\begin{enumerate}[label=A\arabic*)]
    %\item $\left<n\right|O\left|n\right>$ changes very little between eigenstates with similar energy, 
    %\item the level spacing, $E_n-E_{n+1}$, is sufficiently small, 
    %\item and the energy uncertainty of the initial state is sufficiently small,
%\end{enumerate}
then an eigenstate, randomly selected from a micro-canonical ensemble at the energy of the initial state, will describe the long time average observable (LTO): $O(t,T)\approx\left<n\right|O\left|n\right>$ for large $t$ and $T$.

{ ETH was originally discussed\cite{deutsch1991,srednicki1994} in classically chaotic systems where the eigenstates behave similar to random matrices and allows one to hypothesize additional structure on the off diagonal matrix elements of observables, $\left<n\right|O\left|m\right>$.
While this stronger version of ETH allows one to predict relaxation times and response functions\cite{dalessio2016a}, we will {focus on an integrable model and therefore restrict our attention to the weaker version presented above and questions regarding the memory apparent in long time averages.}
%, thermalization is described as a process in which a larger subsystem acts as a thermal bath for a smaller one, and the reduced density matrix of the smaller system is approximated as a Gibbs ensemble\cite{rigol2008}.

In extended systems, the standard mechanism for the breakdown of thermalization is the emergence of an extensive set of conserved charges due to underlying integrability\cite{polkovnikov2011a,batchelor2016,cassidy2011} or a random disorder potential\cite{imbrie2017,nandkishore2015,abanin2019}.
In few mode bosonic systems, thermalization has been predicted from semi-classical chaos\cite{tikhonenkov2013,arwas2015a,khripkov2016a,khripkov2018a,arwas2016,arwas2017a,arwas2019a, PiziiFixPOints}, and it was recently shown that thermalization could fail when an oscillatory drive produced a time crystal\cite{PizziTimeCrystal}.
%In few mode bosonic systems, micro-canonical thermalization has been investigated using a semi-classical analysis\cite{tikhonenkov2013,arwas2015a,khripkov2016a,khripkov2018a,arwas2016,arwas2017a,arwas2019a, PiziiFixPOints}.
%In these works, the loss of initial state memory is explained by the appearance of chaos, and it was recently shown that thermalization could fail in these systems when an oscillatory drive produced a time crystal\cite{PizziTimeCrystal}.

In this article, we explore a similar phenomenon for the long time behavior of a quantum evolution, but for a system which is not extended nor classically chaotic.
The model we study is that of an SU(2) spin with large fixed size $|J|>50$ and evolving with respect to the Hamiltonian 
\begin{eqnarray}
    H=-J_x+\frac{\Lambda}{2|J|}J_z^2 ,
\end{eqnarray}
where $J_x, J_z$ and $J_y$ are the canonical SU(2) spin operators, and we assume $\Lambda>1$.
We formulate the question of thermalization for this system by asking: 1) for which initial states do LTOs thermalize and approach a micro-canonical ensemble, and 2) for states that do not thermalize, what is the mechanism that maintains information about the initial state.
{ We focus our analysis on the time averages, $T>>1$, of observables $O=J_x$ and $O=J_z$, but check by exact calculation that the results remain unchanged for $T\rightarrow 0$.}

\begin{figure}
    \includegraphics[width=0.4\textwidth]{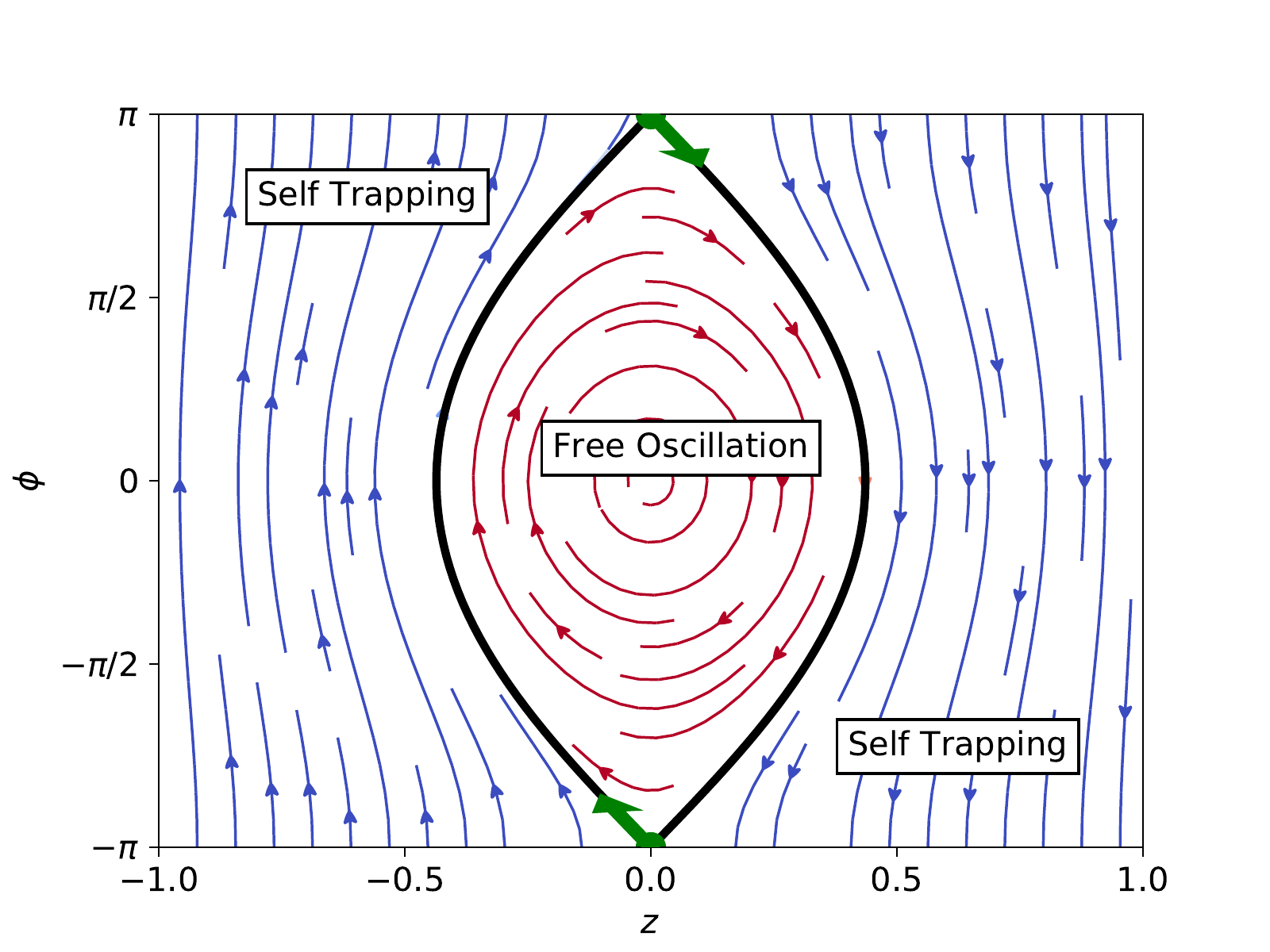}
    \caption{
        Classical trajectories: the separatrix is shown in black (bold) and separates the circular free-oscillation trajectories (red) from the self-trapping ones (blue).
        The red dots mark the unstable fixed point at $(z=0,\phi=\pm\pi)$ and the green arrows mark the unstable directions.
    }
    \label{fig:classicalFlowA}
\end{figure}
This spin Hamiltonian is expected to describe boson tunneling experiments\cite{zibold2010,strobel2014}, and the theory community has explored its dynamics\cite{raghavan1999,micheli2003,mahmud2005,chuchem2010,huang2012,lapert2012,khripkov2013,lovas2017,mathew2017, kelly2019,hennig20212,MORITA2006337}.
{
    While expressions for exact eigenstates\cite{MORITA2006337} do not transparently answer the above questions,
we find particularly useful a semi-classical analysis\cite{raghavan1999,micheli2003,mahmud2005,polkovnikov2009,chuchem2010,khripkov2013,lovas2017,mathew2017, Mori2017, hennig20212} that} describe the classical trajectories shown in Fig.~\ref{fig:classicalFlowA}.
These trajectories, and corresponding eigenstates, have two distinct behaviors known as Josephson oscillation and self trapping, and are separated by a separatrix at $E=1$.
Unlike the few-mode boson models, these trajectories are not chaotic and relaxation occurs through quantum effects\cite{khripkov2013}.
Thus, to consider the question of thermalization, we use the correspondence between classical trajectories and eigenstates {given by the WKB quantization procedure} to access the assumptions required by ETH and answer the questions posed above.
{We first find that for initial states with energy sufficiently different from the energy of the separatrix, $E=1$, { the assumptions A1,A2 and A3 of ETH are obeyed (similar to results in \cite{Mori2017}) and observables come to an equilibrium described by a micro-canonical ensemble}.}

{The primary result of this work finds that, for initial states with energy on the separatrix, the assumptions of ETH do not hold and LTO do not thermalize.
Perhaps most surprisingly, thermalization is avoided by a memory mechanism that remembers a quantity not conserved by the classical dynamics, the initial phase $\phi$.}  
We find that this memory can be explained by a set of eigenstates becoming localized around the unstable fixed point shown in Fig.~\ref{fig:classicalFlowA}.
{This localization was first observed by \cite{chuchem2010}}, but its consequences for the long time memory of initial properties was not investigated.
Elaborating on the analysis developed in \cite{chuchem2010,mathew2017}, we then explain how the localization is due to the asymptotically slow classical dynamics near the unstable fixed point and derive how the long time memory depends on the size of the spin $|J|$. 
We conclude with a discussion on relevant experiments and propose that this mechanism for the breakdown of thermalization is a general phenomenon present in other models which show unstable fixed points in the classical limit.

\begin{figure}
    \includegraphics[width=4cm, height=4cm]{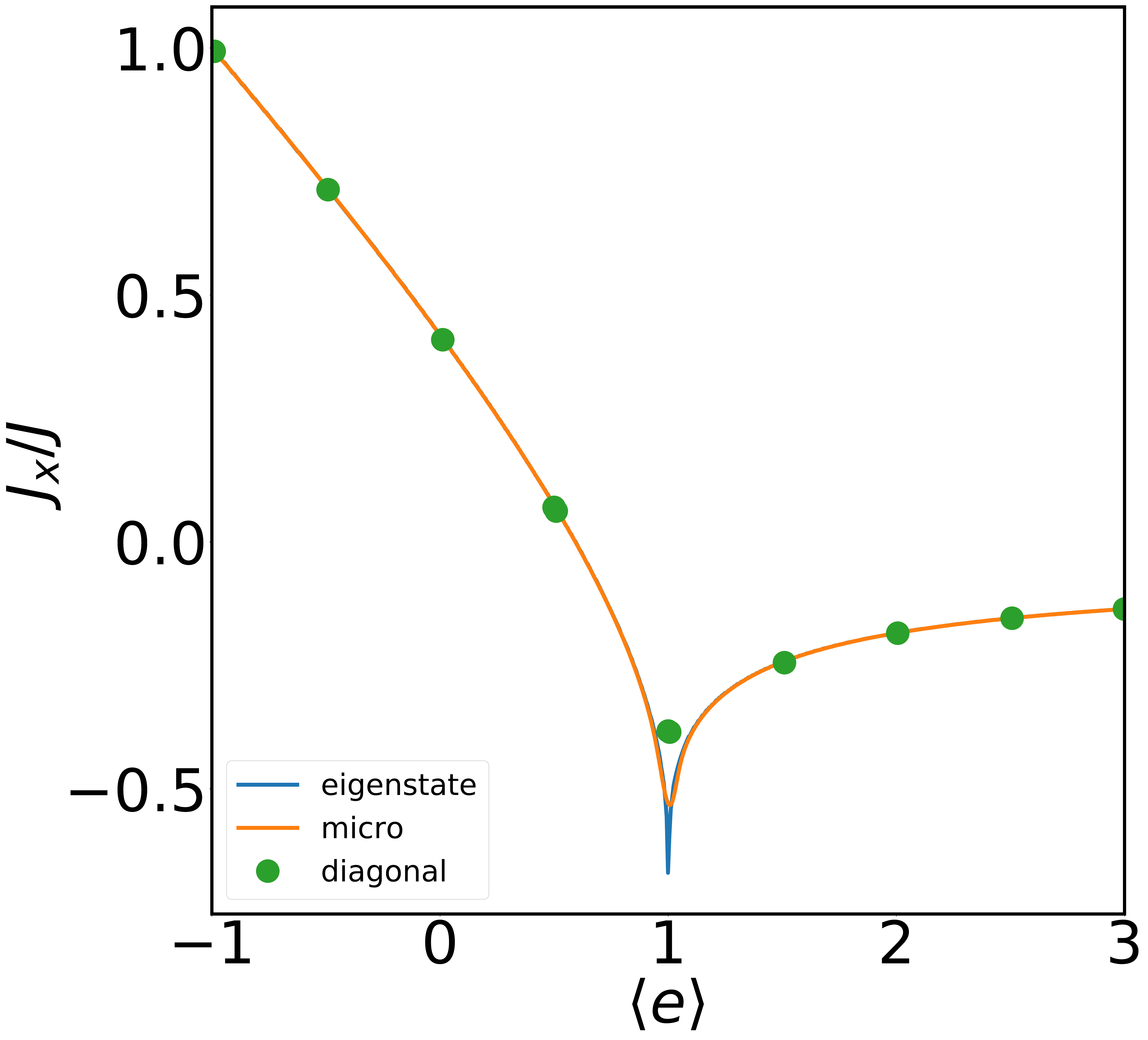}
    \includegraphics[width=4cm, height=4cm]{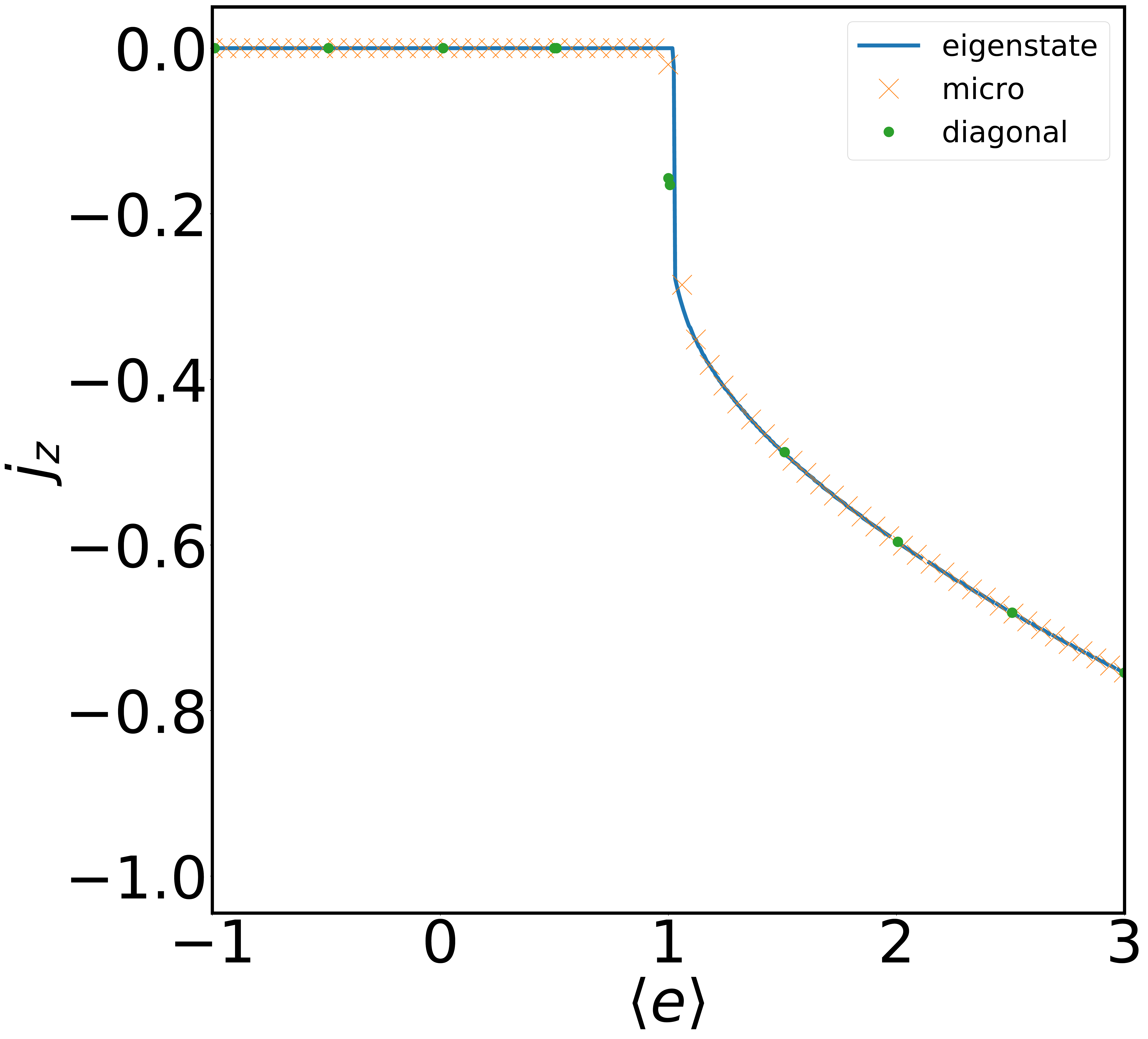}
    \caption{
        (Color online) Energy dependence of ensemble expectations values of $J_x$ and $J_z$:
        The orange line is for the Micro-canonical, while the green dots are for the diagonal ensembles with initial phase $\phi'=0$ and increasing $\left<J_z\right>/J=z'(E)$.
        The energy dependence of the eigenstate expectation values is also shown.
        These calculations where performed for $|J|=1000$ and $\Lambda=10$.
        Notice the strong departure of the diagonal ensemble at $E=1$ from the micro-canonical ensemble.
    }
    \label{fig:classicalFlowB}
\end{figure}
\section{Semi-Classical Picture and ETH:}
We first consider the case when the assumptions of ETH are valid and the large spin thermalizes.
To do so it will be useful to first consider why assumptions of ETH generally imply thermalization.
First consider the eigenstate decomposition of the initial state density matrix, $\sum_{nm}c_{n}c_{m}\left|n\right>\left<m\right|$.
At long times, $t$ and for sufficiently large $T$, we can expect that only the diagonal terms of the density matrix contribute to observables\cite{dalessio2016a,khripkov2013}:
\begin{eqnarray}
    O(t,T)\approx\sum_{m}\left|c_{m}\right|^2\left<m\right|O\left|m\right>.
    \label{eq:energyAverage}
\end{eqnarray}
If A3) of ETH is true, then $\left|c_{m}\right|^2$ is non zero only in a small energy window.
Furthermore, if A1) of ETH holds, then $\left<m\right|O\left|m\right>$ is approximately constant over the eigenstates with significant probability $\left|c_{m}\right|^2$.
Finally A2) ensures there are multiple eigenstates in the micro-canonical ensemble which can be sampled, and we can conclude that a representative eigenstate $\left<n\right|O\left|n\right>$ can be chosen to factor out of the average in Eq.~\ref{eq:energyAverage} yielding: $O(t,T)\approx\left<n\right|O\left|n\right>$.

We now use a semi-classical analysis to determine when these three assumptions of ETH hold for the nonlinear spin Hamiltonian.
The semi-classical analysis is based on a Wigner-function formalism in which states and operators are represented as functions, $W(z,\phi)$ and $O(z,\phi)$, of $z$, the eigenvalue of $J_z/\left|J\right|$, and its conjugate momentum $\phi$. 
In this formalism, the observables $J_z$ and $J_x$ are given by $|J|z$ and $|J|\sqrt{1-z^2}\cos(\phi)$ respectively, and the Hamiltonian is written as\cite{raghavan1999}:
\begin{eqnarray}
    \frac{H(z,\phi)}{|J|}=\frac{\Lambda}{2}z^2-\sqrt{1-z^2}\cos(\phi)
    \label{eq:classicalHamiltonian}
\end{eqnarray}
The expectation values of a state $W(z,\phi)$ with an observable $O(z,\phi)$ is computed with:
\begin{eqnarray}
    \left<\psi\right|O\left|\psi\right>=\frac{1}{4\pi}\int_{-1}^{1}\int_{-\pi}^{\pi}dzd\phi W(z,\phi)O(z,\phi).
    \label{eq:semiObserve}
\end{eqnarray}

We use the set of spin coherent states as our initial states because they are regularly created in experiments\cite{zibold2010,strobel2014}.
In the Wigner-function formalism these states are represented by Gaussian distributions that become more localized around a mean $z'$ and a mean $\phi'$ as the size of the spin, $|J|$, is increased .
Since a state which is more local around a specific $z'$ and $\phi'$ has smaller energy uncertainty, assumption A3) of ETH is satisfied when $|J|$ is sufficiently large.
 
\begin{figure}
    \includegraphics[width=0.23\textwidth]{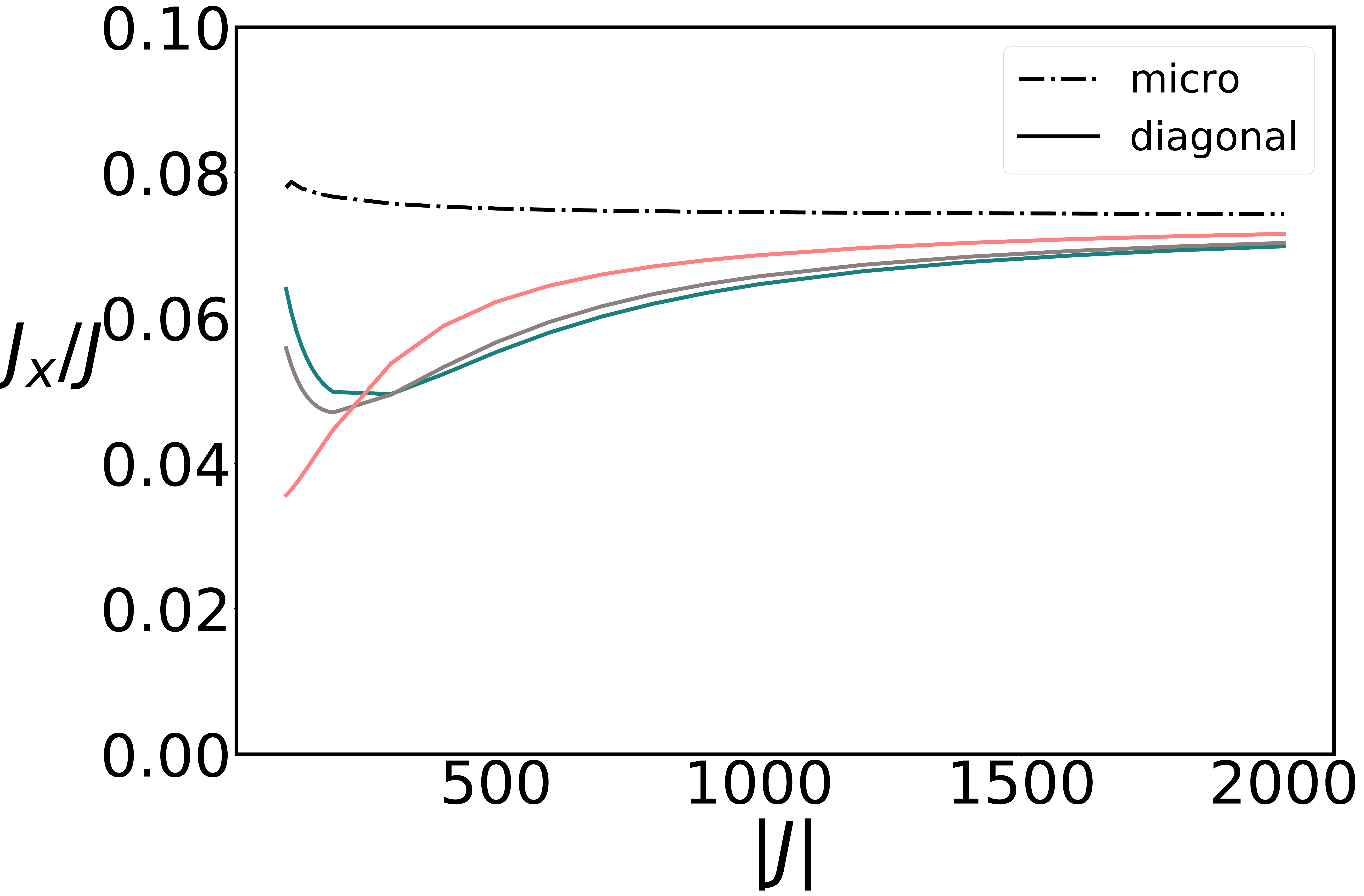}
    \includegraphics[width=0.23\textwidth]{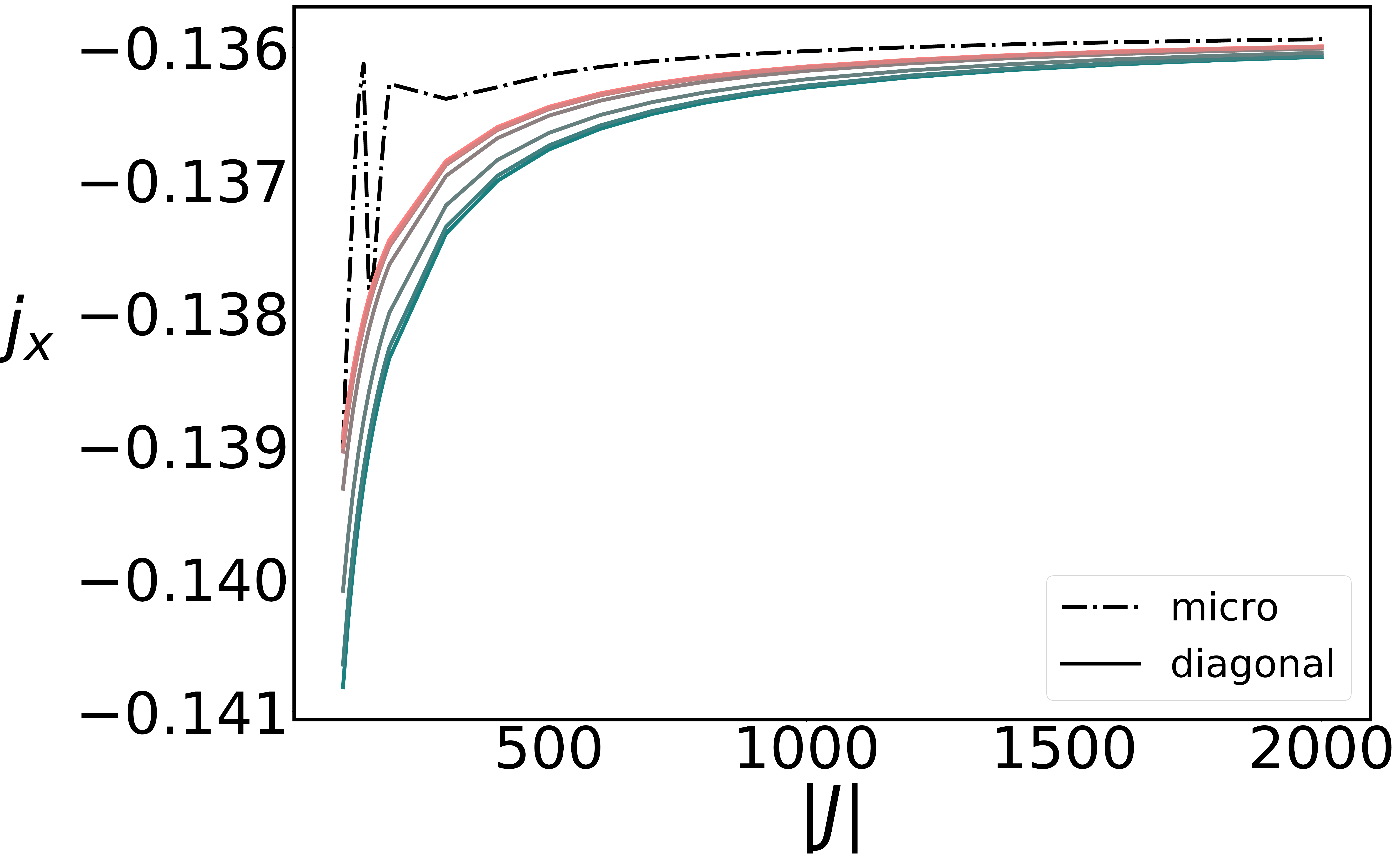}
    \includegraphics[width=0.23\textwidth]{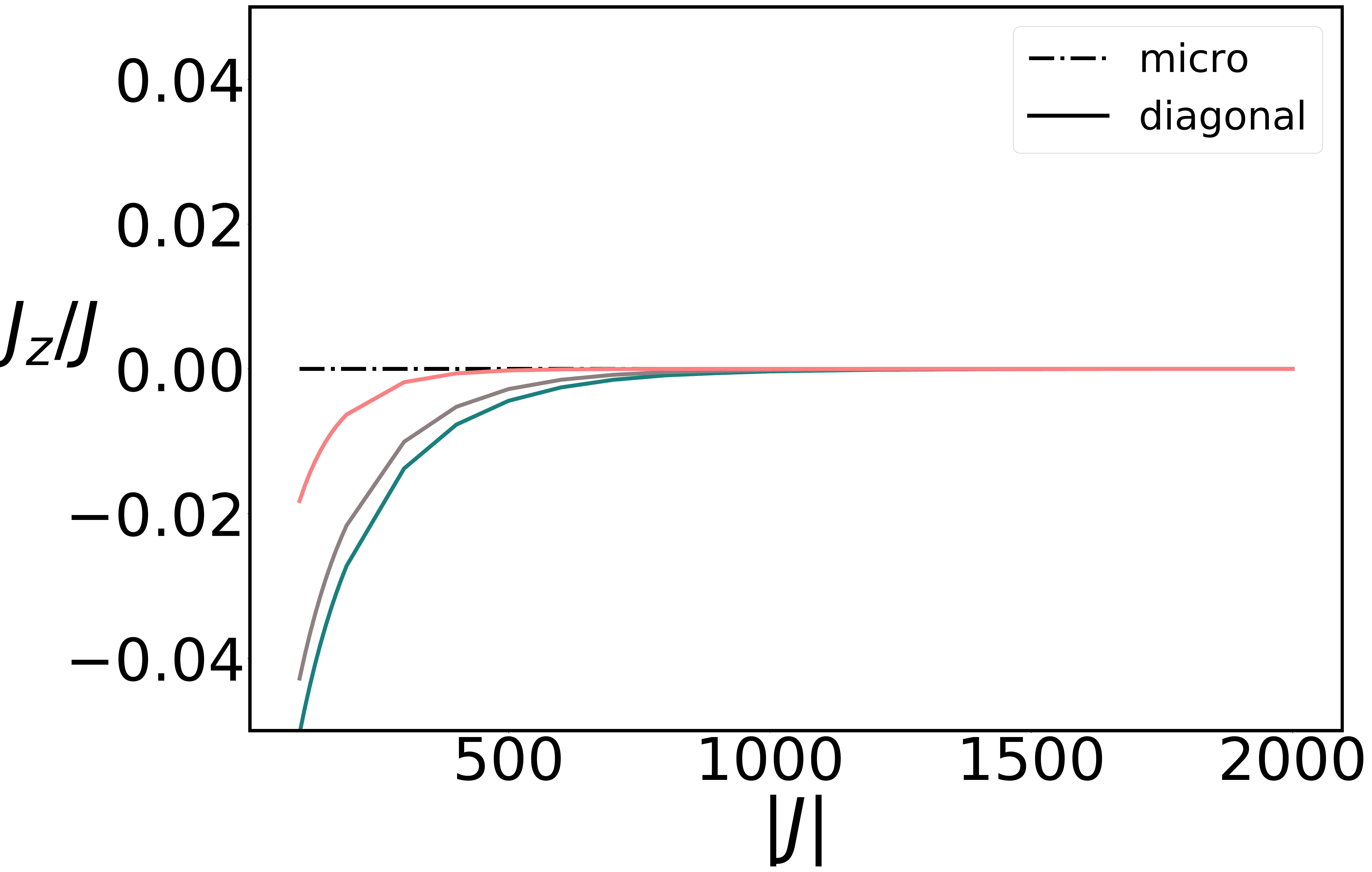}
    \includegraphics[width=0.23\textwidth]{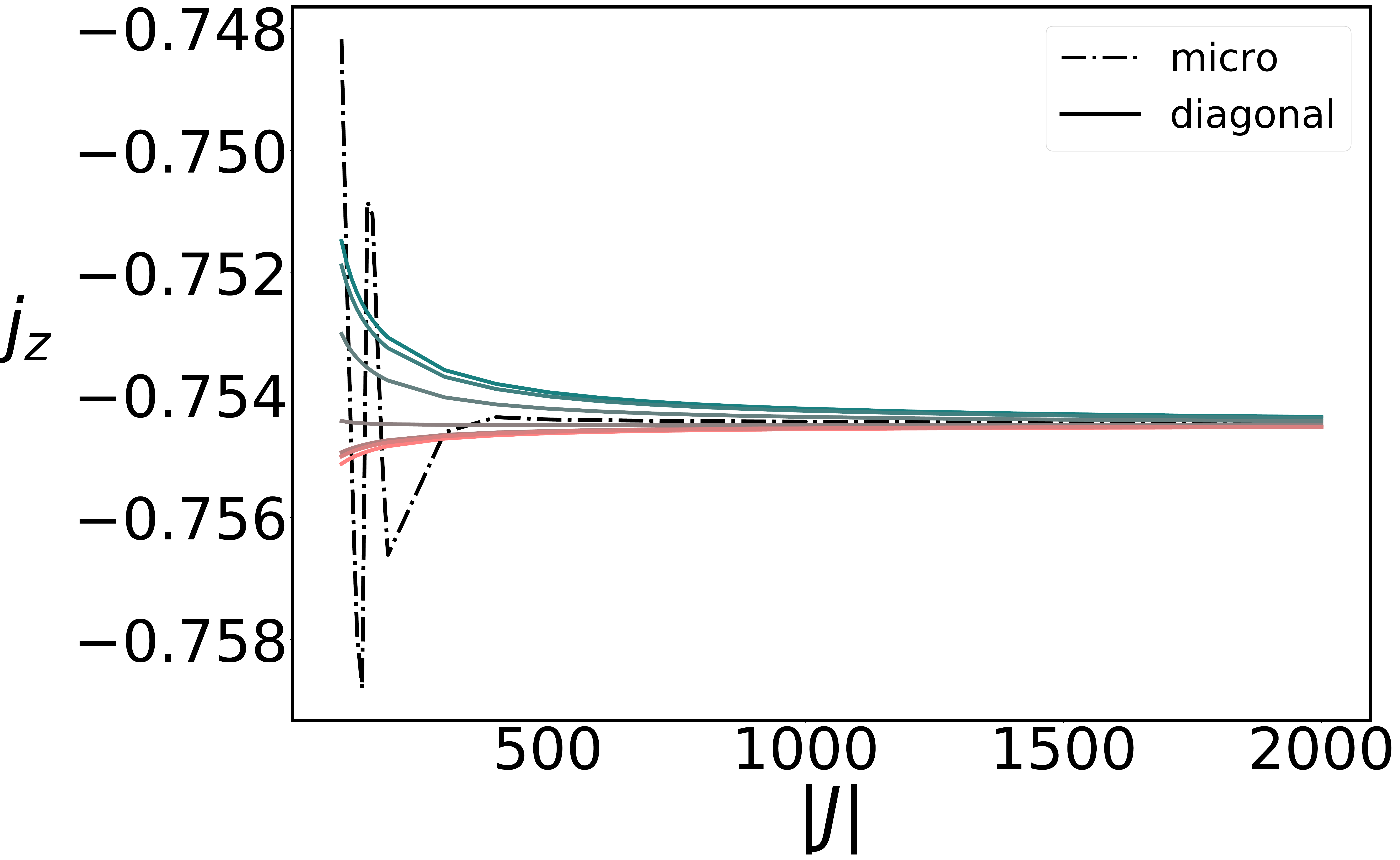}
\caption{(color online) Thermalization for Self-Trapping and Free-Oscillating Dynamics: In this plot, we show the  micro-canonical and diagonal ensemble expectation values as a function of $|J|$ and $\phi'$ for $E=0.5$(1st column) and  $E=3$ (second column).  The color indicates the initial phase $\phi'$ where it ranges from 0(dark blue) to $\pi$ (bright pink).  When $|J|$ increases, the energy level spacing decreases and the assumptions A2) and A3) of ETH become more valid.  Thus for large $|J|$ ETH for the diagonal matrix elements is valid and the dependence of the LTOs on the initial phase is lost. These calculations where done with $\Lambda=10$}
    \label{fig:bareVnA}
\end{figure}

We now consider when assumptions A1) and A2) hold by constructing the Wigner functions of the eigenstates via a semi-classical analysis.
The zeroth order classical analysis treats Eq.~\ref{eq:classicalHamiltonian} as a classical Hamiltonian which yields the periodic trajectories depicted in Fig.~\ref{fig:classicalFlowA}.
Fig.~\ref{fig:classicalFlowA} shows two distinct types of periodic trajectories depending on the energy: for $E<1$, the trajectories known as Josephson oscillation\cite{raghavan1999} occur in which $z$ and $\phi$ periodically oscillates around a stable fixed point at $(z,\phi)=(0,0)$, while for $E>1$ trajectories called self trapping\cite{albiez2005} occur in which $z$ does not change sign, and $\phi$ monotonically increases ($z<0$) or decreases ($z>0$) depending on the sign of $z$.
At $E=1$, there is a separatrix separating the two dynamical behaviors.

Using the correspondence between classical periodic trajectories and eigenstates\cite{sakurai1994modern}, the eigenstate Wigner-functions (EWF) with energy $E$ can be written as {$\rho_{E}(z,\phi)=w(E)\delta \boldsymbol{(}H(z,\phi)-E\left|J\right|\boldsymbol{)}$}, where $w(E)$ is the normalization of the eigenstate with energy $E$.
The quantized energy levels, $E=E_n$, are then determined by the rule\cite{chuchem2010} stating that the area swept out by the eigenstate trajectories is quantized to $2\pi/\left|J\right|$.
Thus, the energy difference between the eigenstate trajectories goes to $0$ as $|J|$ is increased, and assumption A2) of ETH holds true.

Considering assumption A1), we first identify that the Hamiltonian in Eq.~\ref{eq:classicalHamiltonian} has two distinct types of eigenstates corresponding to the Josephson oscillation and the self trapping trajectories.
The self trapping eigenstates are further structured because, for a given energy $E>1$, there are two disconnected trajectories depending on the initial sign of $z$.
These two trajectories will be identified with the sign of $z$ and their associated EWFs are calculated by selecting the correct trajectory when inverting $H(z,\phi)$:
\begin{eqnarray}
    \rho_{E\pm}(z,\phi)=w(E)\left|\frac{dH(z,\phi)}{dz}\right|^{-1}\delta\boldsymbol{(}z\pm\left|H^{-1}(E,\phi)\right|\boldsymbol{)}
\label{eq:semiEigen}
\end{eqnarray}
At lowest order in a semi-classical expansion, these two trajectories correspond to two degenerate eigenstates, while at higher order the degeneracy is lifted\cite{pudlik2014} with splitting exponentially decreasing with $\left|J\right|$.
Since this splitting is exponentially small, we will ignore it and assume all measurements occur before its dynamics are realized( $t<T_{t}\approx e^{|J|}$). 

{
For $E\neq1$, the eigenstate observables will be smooth in energy because, the difference between two neighboring eigenstate trajectories decreases to $0$ as $|J|$ is increased.
While for $E=1$, the self trapping trajectories meet the free oscillating ones, a discontinuity emerges, and non analytic behavior of the eigenstate observables is expected.
The behavior of the eigenstate observables has been identified previously\cite{raghavan1999,benitez2009} and we confirm for $J_x$ and $J_z$ in Fig.~\ref{fig:classicalFlowB}.}

Thus, we find that away from $E=1$ and for large enough $\left|J\right|$, the assumptions of ETH hold, and we expect the LTOs to be described by a micro-canonical ensemble. While for eigenstates with energy $E\approx1$, assumption A1) of ETH does not hold, and additional consideration is required to understand the long time behavior. 

\section{Numerical Analysis of the Diagonal Ensemble:}
From the analysis of the previous section we expect initial coherent states with $z'$ and $\phi'$ away from the separatrix to show thermal behavior at long times.
{
Using exact diagonalization, we confirm that memory of the initial state is lost for $E\neq1$.
This is shown in Fig.~\ref{fig:bareVnA}, which demonstrates that the diagonal ensemble for states with different $\phi'$, but same $E$, all reproduce the same LTO.
We also confirm that a micro-canonical ensemble, and a characteristic eigenstate, describe the LTOs.
This is shown in Fig.~\ref{fig:classicalFlowB} for $J_x$ and $J_z$.
}

Since the hamiltonian is integrable, the off diagonal matrix elements are not random as proposed by ETH, and one can not use ETH to argue that the LTO relax to the time averages predicted by the Diagonal ensemble.
Instead, we must check by exact numerical simulation.
Doing so for $J_x$, we find that the self trapping dynamics and free-oscillating dynamics do in fact relax to a constant value independent of the initial phase $\phi'$.
This is shown in Fig.~\ref{fig:dyno}.

Close to $E=1$, the micro-canonical ensemble and the characteristic eigenstate no longer match LTOs.
Failure of the initial states at $E=1$ to thermalize is further demonstrated in Fig.~\ref{fig:bareVnB}, which shows a dramatic dependence of the LTOs on the initial phase, $\phi'$.
This does not invalidate ETH because assumption A1) of ETH does not hold for these eigenstates.
\begin{figure}[h!]
    \includegraphics[width=0.23\textwidth,trim={0 0.2cm 0 1.5cm}]{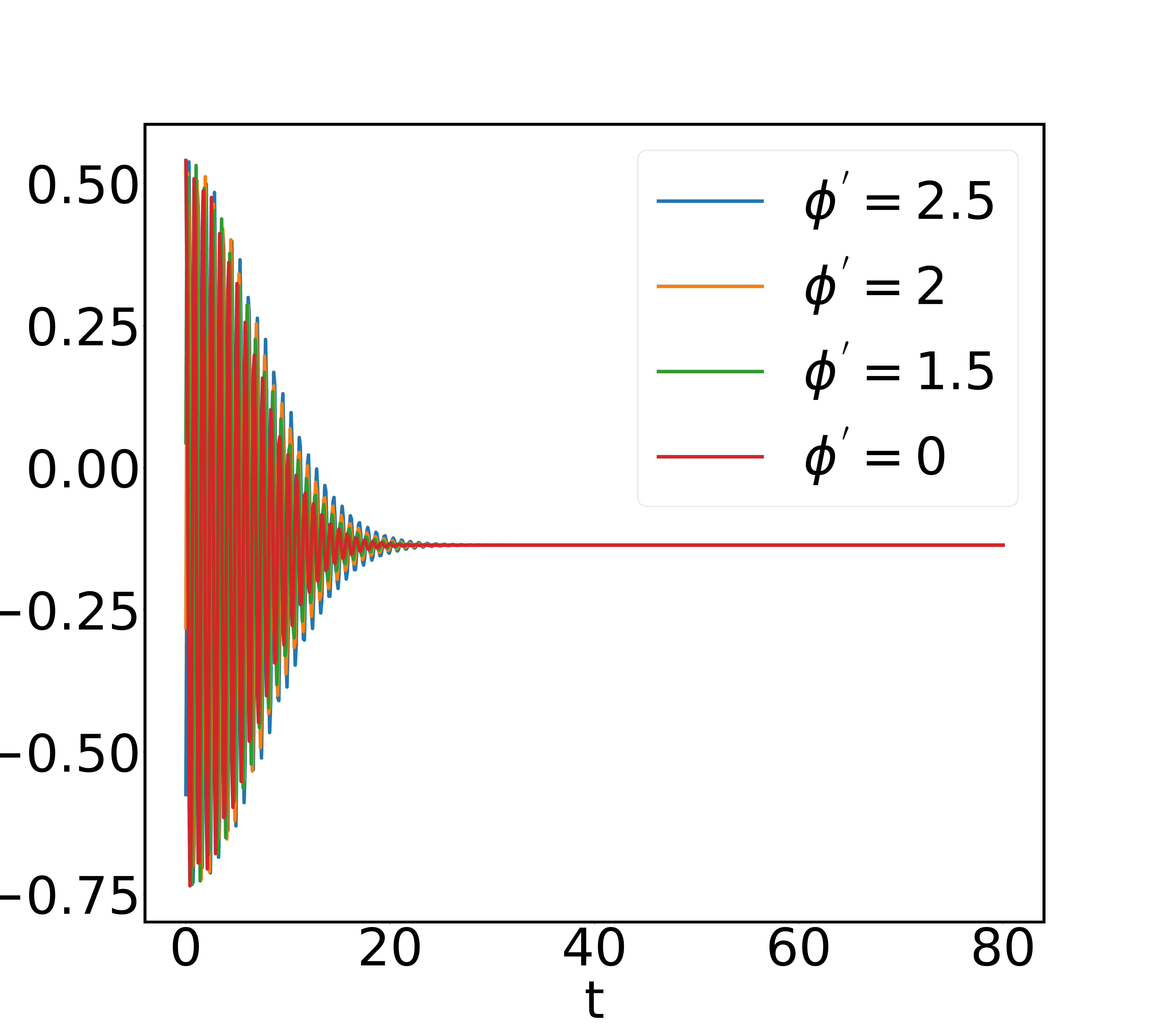}
    \includegraphics[width=0.23\textwidth,trim={0 0 0 0.2cm}]{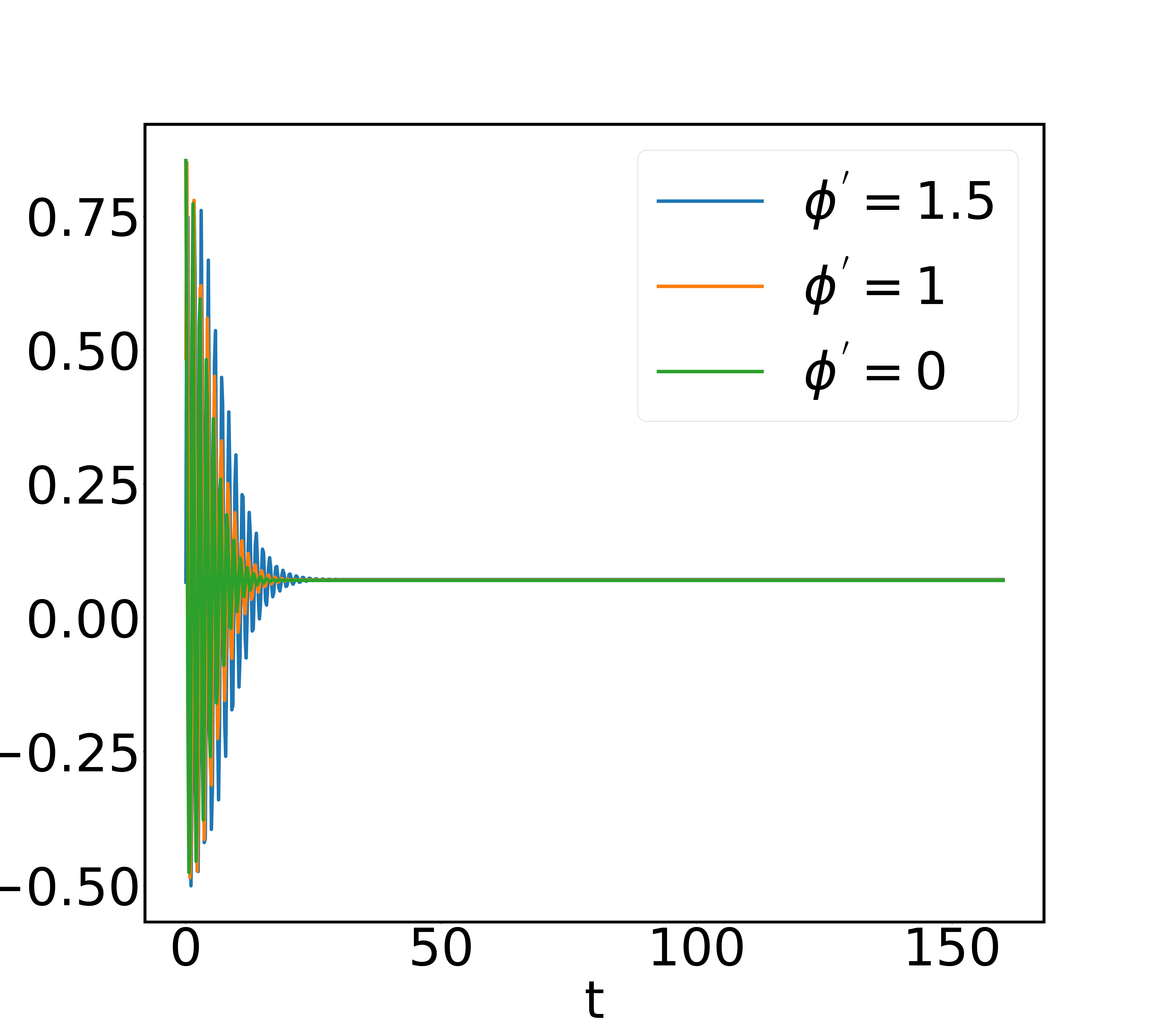}%
    \caption{(Color Online) Dynamics of $J_x/|J|$ for different values of $\phi'$ at energies $E=3$(top), $E=0.5$ (bottom).}
    \label{fig:dyno}
 \end{figure}
 
\begin{figure}
    \includegraphics[width=0.4\textwidth]{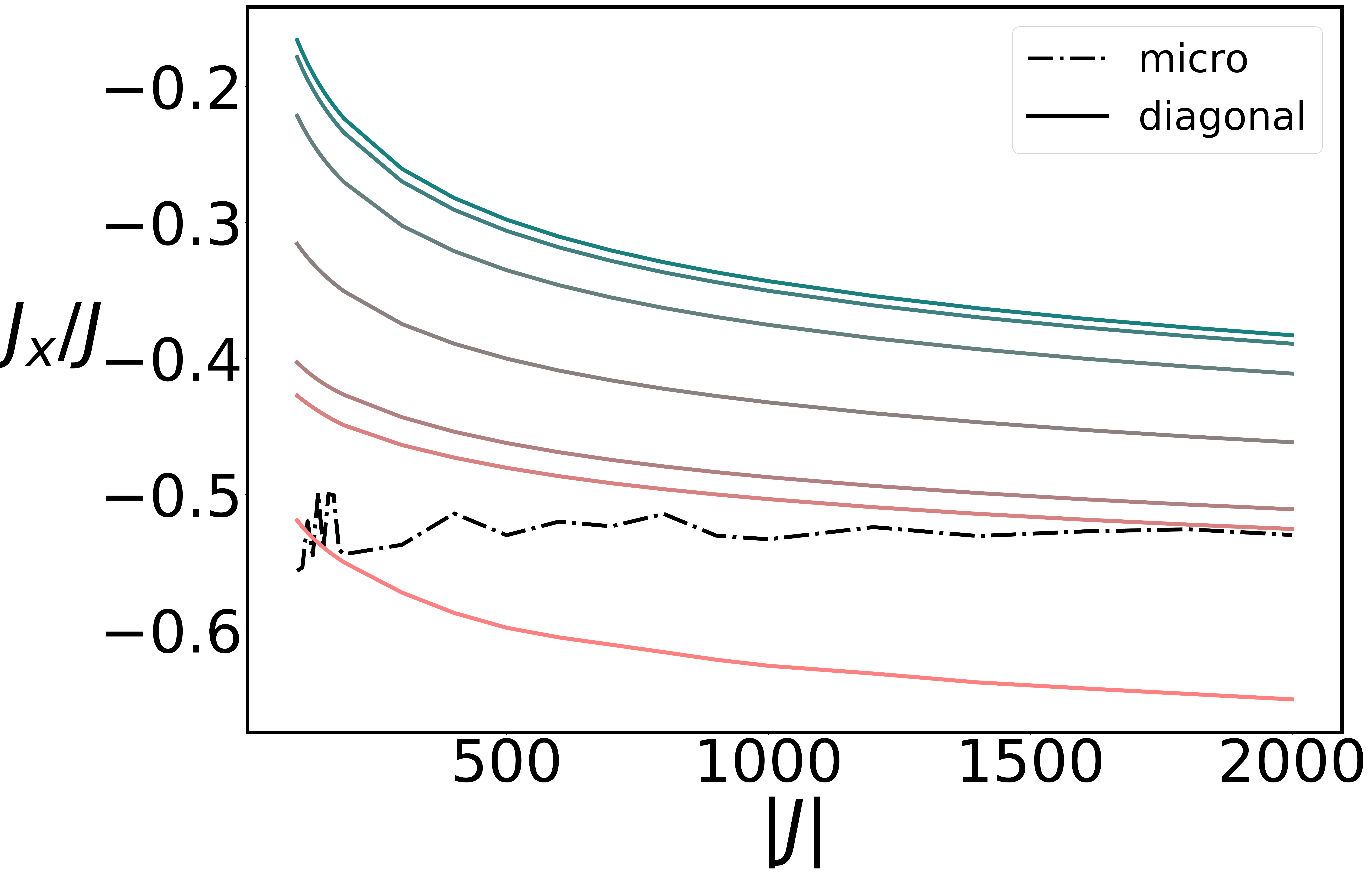}
    \includegraphics[width=0.4\textwidth]{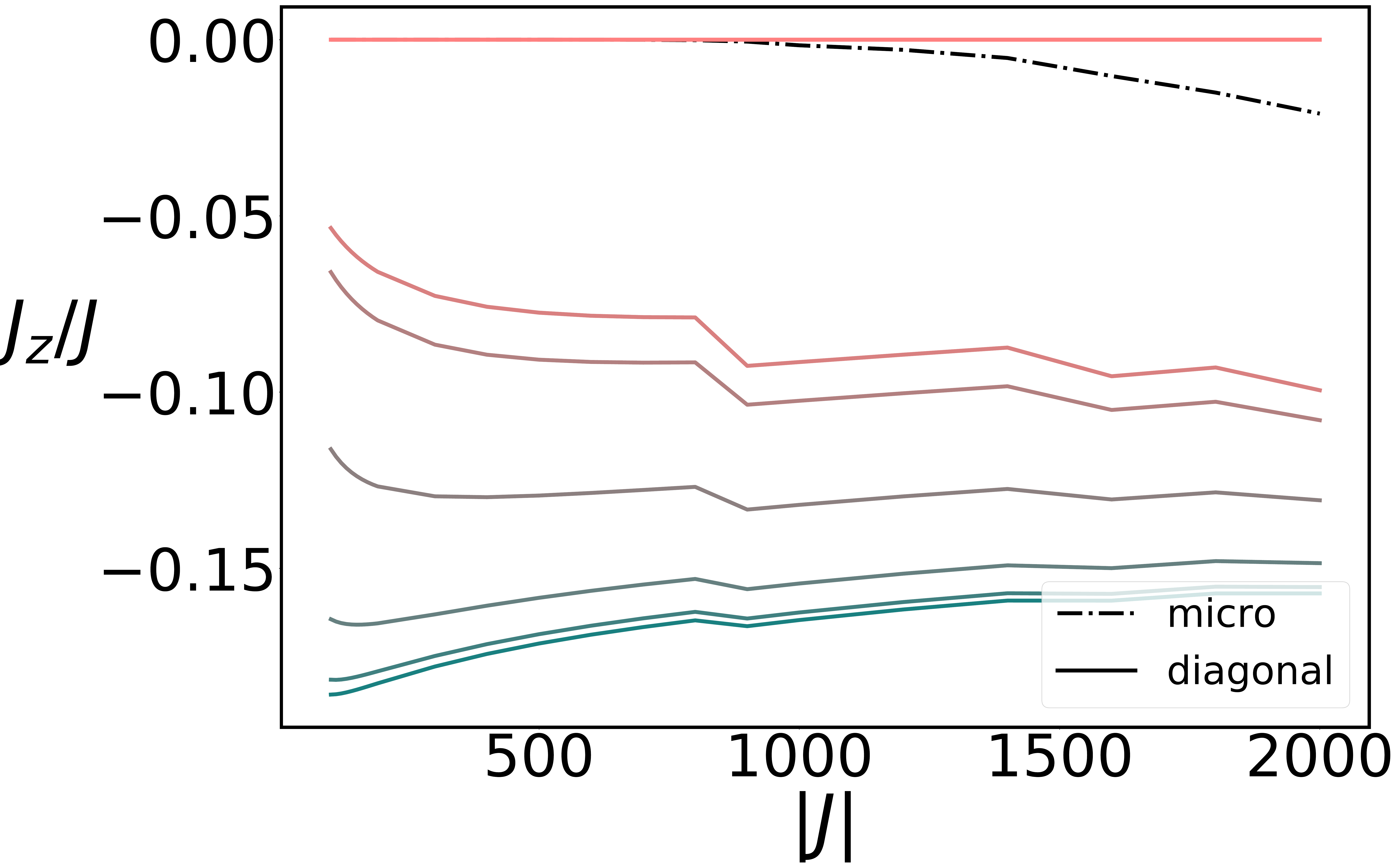}
\caption{(color online) Failure of Thermalization on the Separatrix: The contents of these plots are equivalent to those in Fig.~\ref{fig:bareVnA}, except they are calculated for initial states with $E=1$.  This time, despite the assumptions A2) and A3) becoming more valid as $|J|$ is increased, the assumption A1) remains invalid and the memory of initial phase $\phi'$ remains at large $|J|$.}
    \label{fig:bareVnB}
\end{figure}

\section{Semi-Classical Analysis of the Breakdown of Thermalization:}

To better understand this breakdown of thermalization we investigate, using the semi-classical analysis, how the $E\approx1$ eigenstates affect the LTO of the initial coherent states with $E\approx1$ .
We begin by calculating the diagonal ensemble and its expectation values for the initial coherent states used above.
Semi-classically\cite{chuchem2010} the diagonal ensemble is given as:
\begin{eqnarray}
    \rho_{diag}=\frac{1}{4\pi}\int_{-1}^{1}\int_{-\pi}^{\pi}dzd\phi W_c(z,\phi,z',\phi')\rho_{E,s}(z,\phi)
    \label{eq:eOlap}
\end{eqnarray}
where $W_c$ is the initial coherent state Gaussian distribution centered around $z'$ and $\phi'$ with variance $\sim\frac{1}{J}$, and the EWF, $\rho_{E,\pm}$, is given by the delta function in Eq.~\ref{eq:semiEigen}.

To calculate the LTOs, one must convolve the diagonal ensemble with the eigenstate expectation values:
\begin{eqnarray}
   %& O_{diag}(\phi',z')=&\\ \nonumber
   %& \int_{-1}^{\Lambda/2}dE \sum_{s}\rho_{diag}&(\phi',z',E,s)O(E,s)
  O_{diag}(\phi',z')=
  \int_{-1}^{\Lambda/2}dE \sum_{s}\rho_{diag}(\phi',z',E,s)O(E,s)
    \label{eq:saddleInt}
\end{eqnarray}
where the sum over $s$ is the sum over self trapping states when $E>1$ and a fixed $s=0$ for $E<1$, and $O(E,s)$ is the eigenstate expectation value calculated using Eq.~\ref{eq:semiObserve} with $W(z,\phi)=\rho_{E,s}(z,\phi)$.

Understanding this integral, and consequently why the LTOs encode information about the initial phase $\phi'$, requires understanding the structure of the eigenstates and their EWFs.
While an EWF is constrained to an equal energy surface, the shape of the energy surface affects how the EWF is distributed within the energy surface. %phase space($z$,$\phi$).
This is captured by the Jacobian, $\left|\frac{dH(z,\phi)}{dz}\right|$, which appears in Eq.~\ref{eq:semiEigen} due to the transformation of the energy delta function to phase space coordinates.
Take the $s=1$ self trapping eigenstate for example.
If one integrates out $z$ using the delta function, the Jacobian $|\frac{dH}{dz}|(E,\phi)=|\frac{d\phi}{dt}|(E,\phi)$  weighs the EWF.
Therefore, the EWF will have more weight in regions where $\phi$ is changing slower in time.

On the separatrix, $E=1$, the classical spin comes to a complete stop on the unstable fixed point; the Jacobian limits to $0$, $\text{lim}_{E\rightarrow1}\text{lim}_{\phi\rightarrow\pi}|\frac{dH}{dz}|(E,\phi)=0$; and the EWFs with $E\rightarrow1$ become localized on the unstable fixed point:$\rho_{E\rightarrow1}(z,\phi)\approx\delta(z)\delta(\phi-\pi)$.  
The singularity of this localization result in the non-analytic behavior of the eigenstate expectation values near $E=1$ (see for example $J_x$ in Fig.~\ref{fig:classicalFlowB}).

This singular localization also produces a non-analyticity in the eigenstate overlaps for the set of initial coherent states with $E\approx1$, but $\phi'\neq\pi$.
Since these initial states have Wigner functions localized around $\phi'$ and $z'=H^{-1}(E=1,\phi')$ and the EWFs for $E\approx1$ are localized around $\phi=\pi\neq\phi'$ and $z=0\neq z'$, their overlap integrals in Eq.~\ref{eq:saddleInt} will vanish.

These two non-analyticities are integrated over in Eq.~\ref{eq:saddleInt} and results in the memory effects depicted in Fig.~\ref{fig:bareVnB}.
In one limit, an initial coherent state with $\phi'\approx \pi$ will overlap the unstable fixed point eigenstate at $E=1$, and the LTOs will closely match the observables of that same eigenstate($J_z=0$ and $J_x=-1$).
In the other limit, when the initial $\phi'$ is away from $\pi$, the initial coherent state will have negligible overlap with the $E=1$ eigenstate, the LTOs will depart from the observables of the $E=1$ eigenstate.
This is depicted in Fig.~\ref{fig:bareVnB}, in which the closer $\phi'$ is to $\pi$, the closer $j_z=J_z/J$ and $j_x=J_x/J$ approach $0$ and $-1$ respectively.

\begin{figure}
    \includegraphics[width=0.45\textwidth]{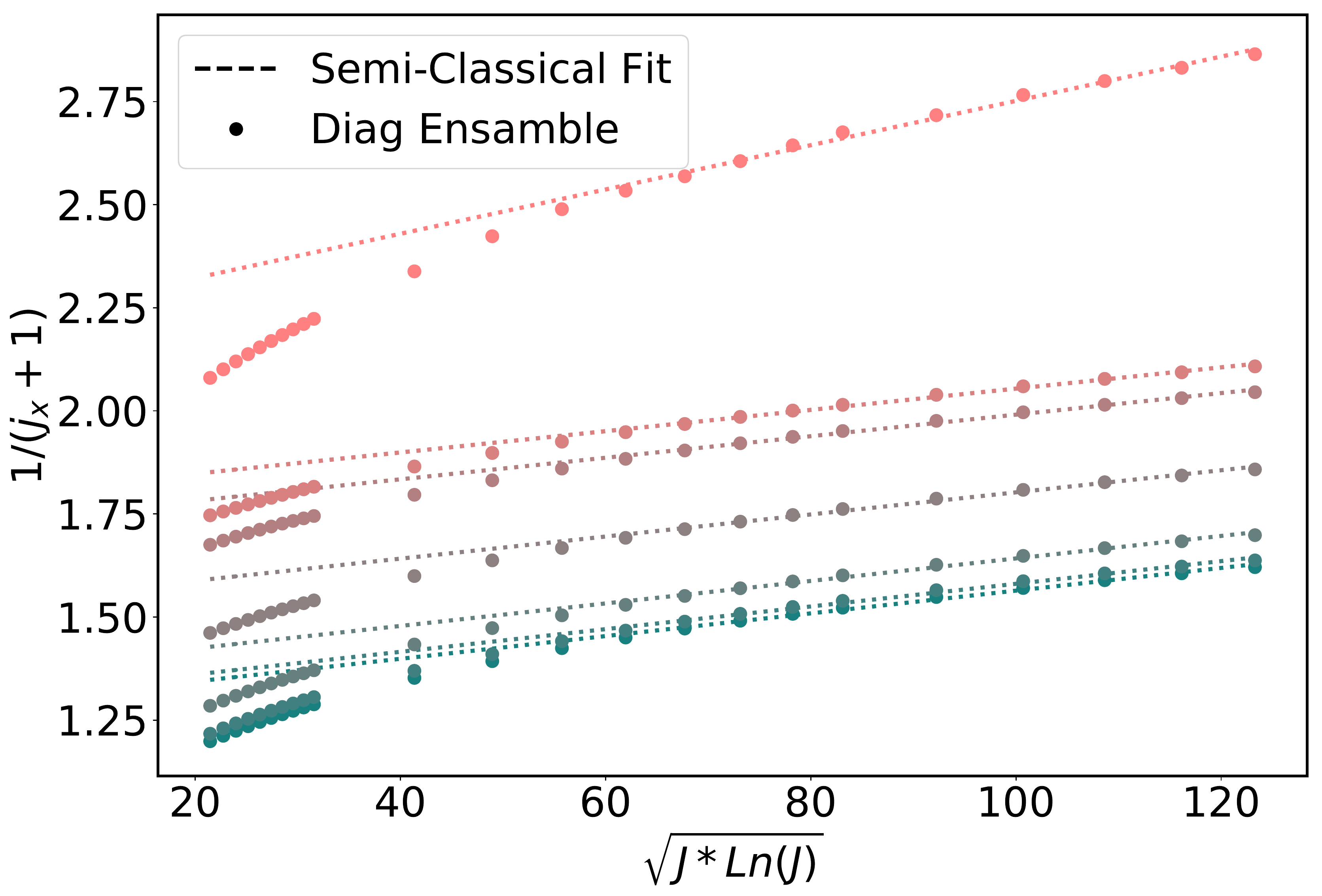}
    \caption{(Color Online) Classical fit to exact numerical calculations of diagonal ensemble. This plot shows the $\sqrt{\left|J\right| ln\left|J\right|}$ scaling of $j_x=J_x/J$ for the diagonal ensemble of a set of initial states with energy $E=1$ and different $\phi'$ and $\left|J\right|$.  The dots are computed using exact diagonalization and the color indicates the initial $\phi'$ as in Fig.~\ref{fig:bareVnA}. Eq.~\ref{eq:cfit} predicts $\sqrt{\left|J\right| ln\left|J\right|}$ and the linear dashed lines are given by Eq.~\ref{eq:cfit} with $F(\phi')$ fit to match the exact calculations for $|J|>500$.}
    \label{fig:semiClassicalCalc}
\end{figure}

\section{Large $|J|$ behaviour of Initial State Memory}
To capture this behavior analytically, we perform a saddle point expansion for the integral Eq.~\ref{eq:saddleInt}.
{A similar saddle point approximation was done in \cite{mathew2017,chuchem2010}, but only for an initial state on the unstable fixed point.
To capture how the long time memory depends on the size of the spin $\left|J\right|$, we perform the saddle point for initial states computed off the unstable fixed point.
The results in \cite{mathew2017,chuchem2010} will not work here because the diagonal ensemble has a qualitatively different saddle point structure for states on and off the unstable fixed points\cite{chuchem2010}.}

To perform the saddle point approximation away from the unstable fixed point, we begin with finding the diagonal ensemble $\rho_{diag}(\phi',z',E,z)$ by evaluating the integral in Eq.~\ref{eq:eOlap}.
For large $|J|$, the integral is restricted over a region in the vicinity of $z'$ and $\phi'$.
Since this region is away from the unstable fixed point, the equal energy countor can be approximated as a line and the Jacobian $\left|\frac{dH(z,\phi)}{dz}\right|^{-1}$ is approximately constant.
Performing the Dirac delta and Gaussian integrations yields:
\begin{eqnarray}
    \rho_{diag}&(\phi',z',E,s)\sim e^{-\frac{(E-H(\phi',z'))^2}{2\sigma^2(\phi',z')}+\text{ln}(w(E))}
        \label{eq:diagclassical}
\end{eqnarray}
where $w(E)$ is the eigenstate normalization.
The Gaussian variance $\sigma(\phi',z')$ is given in Appendix~\ref{apx:sigma}, and scales with $\left|J\right|$ as $\sim\frac{1}{\sqrt{\left|J\right|}}$  with proportionality dependent on $\phi'$ and $z'$.

For initial states away from the separatrix, $w(E)$ is approximately constant\cite{chuchem2010}, and the diagonal ensemble is a gaussian.
This is not the case for initial states on the separatrix.
Instead, the asymptotically slow dynamics, and consequently the asymptotic divergence of the Jacobian, near the unstable fixed point, forces an asymptotic vanishing of $w(E)$, and consequently $\rho_{diag}(\phi',z',E,s)$, at $E=1$.
Computing $w(E)$ in Appendix~\ref{apx:omega}, we find that it vanishes as $\sim 1/Ln(|1-E|)$, where the proportionality is different depending on if $E$ is greater or less then $1$.

Since, $\rho_{diag}(\phi',z',E,s)$ is $0$ at $E=1$ and in the limit $E\rightarrow\pm\infty$, it possesses a double peak structure in $E$.
{This is qualitatively different from the single peak saddle point structure used to perform the calculations in \cite{mathew2017,chuchem2010}.}
The locations of these two peaks determines our saddles and are computed in Appendix~\ref{apx:saddles}.
In the large $|J|$ limit, these saddles become symmetric about $E=1$ given by $E_s=1\pm \delta$, and go to $E=1$ as $ \delta\sim\frac{1}{2JLn(J)}$.

We then evaluate the integral in Eq.~\ref{eq:saddleInt} at these saddles:
\begin{eqnarray}
    &O_{diag}(z',\phi')=\int \rho_{diag}(E,z',\phi')O(E)\approx\\ \nonumber
    &\frac{1}{3}\left[2O(1+|\delta|)+O(1-|\delta|)\right]
\end{eqnarray}
where the factor of $2$ for $+|\delta|$ occurs because $\rho_{diag}(E,z',\phi')$ in the $\delta\rightarrow+0$ limit is twice as large as in the $\delta\rightarrow-0$ limit(See $w(E)$ in Appendix~\ref{apx:omega}).
In Appendix~\ref{apx:obs} we compute $J_x(1\pm|\delta|)$ and $J_z( 1\pm|\delta|)$ for small $\delta$ using methods similar to \cite{mathew2017, chuchem2010}.
Using these results, we get:
\begin{eqnarray}
    \label{eq:cfit}
    j_{z,diag}(\left|J\right|,E=1,\phi')&= &\frac{4\pi\sqrt{(\Lambda-1})}{3\Lambda ln\left[F(\phi')\left|J\right|ln[\left|J\right|]\right]} \\ \nonumber
    j_{x,diag}(\left|J\right|,E=1,\phi')&= &-1+\frac{1}{\sqrt{F(\phi')\left|J\right|ln[\left|J\right|]}}\frac{3+\Lambda}{3(\Lambda-1)}
\end{eqnarray}
where the factor $F(\phi')=\left[2\sigma(z',\phi')^2\left|J\right|\right]^{-1}$, $j_{x(z)}=J_{x(z)}/J$, and $z'$ is fixed by energy $z'=H^{-1}(E=1,\phi')$. 
The factor $F(\phi')$ is constant in $\left|J\right|$ but has a non-trivial dependence on the initial phase $\phi'$ via $\sigma(z',\phi')$, the energy variance of the coherent state.
This non-trivial dependence in $\phi'$ describes the memory effects shown in Fig.~\ref{fig:bareVnB} for the initial states with $\phi'\neq \pi$.
{For the initial states with $\phi'\approx\pi$ we must use the single peak saddle point approximation outlined by \cite{chuchem2010,mathew2017} which give a different scaling to the fixed point values of $J_x\rightarrow -1$ and $J_z=0$.}

While the exact diagonal ensemble for $J_z$ becomes numerically unstable for large $|J|$, we can still compare  exact results for $J_x$ with Eq.~\ref{eq:cfit}.
This comparison is shown in Fig.~\ref{fig:semiClassicalCalc}, where the $\sqrt{\left|J\right|ln[\left|J\right|]}$ scaling is confirmed.
%{\sout{Note that the breakdown of thermalization is a quantum effect which is lost in the $J\rightarrow \infty$ classical limit.}}

\section{Discussion and Possible Experimental Realizations:}
Above we discussed how, for the large non-linear spin with energy $E\neq1$, the assumptions of ETH hold and the spin thermalizes, while for $E=1$ the spin does not thermalize.
This lack of thermalization is particularly interesting because the remembered quantity, $\phi$, is not a conserved quantity of the integrable classical dynamics.  It is therefore a novel form of quantum memory, which is lost in the classical limit $J\rightarrow\infty$ (See Eq.~\ref{eq:cfit}).

{
    Our results are particularly important in the context of recent works on out-of-time order correlations (OTOCs)\cite{Pilatowsky2020,Tianrui2020,Rozenbaum2020}.
    Recently OTOCs have become a diagnostic of quantum many body chaos, and have been shown to display exponentially fast growth when the dynamics of an effective classical system displays chaos\cite{lashkari2013,maldacena2016a,roberts2015,shenker2014,shenker2014a,shenker2015}.
    In the works \cite{Pilatowsky2020,Tianrui2020,Rozenbaum2020}, they found that classically unstable fixed points can produce exponentially growing OTOCs in systems with an integrable classical counterpart, and suggest exponential growth of OTOCs is not a predictor of quantum chaos\cite{Tianrui2020}.
    Our results further support this conclusion, showing that despite the chaotic like behavior suggested by OTOCs, dynamics near the unstable fixed point are precisely those which depict long time memory of an initial state.
}

The appearance of unstable fixed points in semi-classical dynamics is ubiquitous, and we expect this mechanism for the breakdown of thermalization to be general.
{
    While here we discussed a classically two-dimensional, integrable system, the Berry Conjecture\cite{Berry_1977, dalessio2016a} suggests that the correspondence of eigenstates to trajectories, generalizes to a correspondence to micro-canonical ensembles in higher dimensional chaotic systems.
    Since the micro-canonical ensemble is also described by a delta function in energy, the Jacobian produced when transforming to the phase space coordinates would again reveal localization due to slow classical dynamics.
One might again expect singularities due to a localized eigenstate and for them to produce memory effects following similar arguments as discussed above.  
This time, rather than the phase along a separatrix, it would be the distance to the unstable fixed point on the energy surface that is remembered.
This is an exciting possibility which requires further investigation.}
%Quantum systems with integrable classical dynamics offer the simplest generalization, because the correspondence between classical trajectories and eigenstates still holds and the analysis above can be easily repeated.  
%For classically chaotic systems, the same correspondence does not hold, but the eigenstates must still capture the slow dynamics in the semi-classical limit.
%Thus, it is reasonable to still expect the existence of a set of eigenstates that are localized on the unstable classical fixed points and responsible for the slow dynamics.

This mechanism for the breakdown of thermalization may be observable in ultra cold BECs\cite{zibold2010,muessel2015,muessel2015} in which the bosons can be condensed into one of two modes such as two different hyperfine states.
A spin boson mapping then yields the non-linear spin Hamiltonian, where the parameter $\Lambda$ is a ratio between the bosonic interaction energy and the energy associated with the tunneling between the two modes.
{ Previous work has suggested that the other bosonic modes do not affect the dynamics on experimental time scales\cite{lovas2017,khripkov2013}.
Future work may find it interesting to investigate the effect of additional modes and may find connection with other forms of novel long time dynamics\cite{lerose2018}.}

\begin{acknowledgments}
\textbf{Acknowledgments:} This work was supported in part by the NSF under Grant No. DMR-1411345, S. P. K. acknowledges financial support from the UC Office of the President through the UC Laboratory Fees Research Program, Award Number LGF-17- 476883.
The research of E. T. in the work presented in this manuscript was supported by the Laboratory Directed Research and Development program of Los Alamos National Laboratory under project number 20180045DR.
\end{acknowledgments}

\appendix

\section{Eigenstate normalization: $\omega(E)$}.
\label{apx:omega}
In the main text we defined the semi-classical eigenstate Wigner function (EWF) as:
\begin{eqnarray}
    \rho_{E}(z,\phi)=\omega(E)\delta(H(z,\phi)-E)
\end{eqnarray}
where the Hamiltonian is given as:
\begin{eqnarray}
    H=\frac{\Lambda  z^2}{2}-\sqrt{1-z^2} \cos (\phi ),
\end{eqnarray}
and the normalization $\omega(E)$ is given as:
\begin{eqnarray}
    \omega(E)^{-1}=\int \int dz d\phi \delta (E-H(z,\phi)).
\end{eqnarray}
To compute this integral, we focus on the energy close to the separatrix, $E=1\pm\left|\delta\right|$, and expand the Hamiltonian around $E=1$:
\begin{eqnarray}
    H-1=\frac{\Lambda -1}{2} z^2-\frac{(\phi-\pi) ^2}{2}
\end{eqnarray}
Close to the unstable fixed point the trajectories trace out a hyperbola:
\begin{eqnarray}
    z&=&\pm\sqrt{\frac{2}{\Lambda -1}} \sqrt{\frac{\phi ^2}{2}+(E-1)}\\\nonumber
    \phi&=&\sqrt{2(1-E)+(\Lambda-1)z^{2}}
\end{eqnarray}
The Jacobian for both these trajectories are:
\begin{eqnarray}
    \left|\frac{dH}{dz}\right|&=(\Lambda-1)z=& \sqrt{(\Lambda -1)} \sqrt{2(E-1)+\phi ^2} \\ \nonumber
    \left|\frac{dH}{d\phi}\right|&=\phi=&\sqrt{(\Lambda -1)}\sqrt{\frac{2(1-E)}{(\Lambda-1)}+z^{2}}.
\end{eqnarray}
Since the inverse Jacobians, $\left|\frac{dH}{d\phi}\right|^{-1}$ and $\left|\frac{dH}{dz}\right|^{-1}$, contribute the most near the unstable fixed point and we can expand the integrand for $\omega(E)^{-1}$ near them and write:
\begin{eqnarray}
    \omega(1+|\delta|)^{-1}&=\int_{-r_{+}}^{r_{+}} \left|\frac{dH}{dz}\right|^{-1}(\phi,\delta)+C_{+}\\\nonumber
    \omega(1-|\delta|)^{-1}&=\int_{-r_{-}}^{r_{-}} \left|\frac{dH}{d\phi}\right|^{-1}(z,\delta)+C_{-}
\end{eqnarray}
where $r_{\pm}$ denotes the limits where the hyperbolic expansion is valid and $C_{\pm}$ are small and approximately constant for $\delta$ small.
Defining $a$ as:
\begin{eqnarray}
    a_{+}=2(E-1)\\ \nonumber
    a_{-}=\frac{2(1-E)}{\Lambda-1},
\end{eqnarray}
these integrals can be expressed as:
\begin{eqnarray}
    \frac{1}{\sqrt{a(\Lambda-1})}\int_{-r}^{r} \frac{1}{\sqrt{1-a^{-1}x^{2}}}dx = \\\nonumber
    \frac{1}{\sqrt{(\Lambda-1})}\left[\sinh ^{-1}\left(\frac{r}{\sqrt{a}}\right)\right]
\end{eqnarray}
and for $E\approx 1$, this approximates to as:
\begin{eqnarray}
\omega(1+|\delta|)^{-1}=-\frac{Ln\left(|\delta|\right)}{2\sqrt{(\Lambda-1})} \\ \nonumber
\omega(1-|\delta|)^{-1}=-\frac{Ln\left(|\delta|\right)}{\sqrt{(\Lambda-1})} 
\end{eqnarray}

\section{Energy Uncertainty of Diagonal Ensemble for a Coherent State: $\sigma$}
\label{apx:sigma}
To approximate the eigenstate overlap for initial states on the separatrix but away from the fixed points, we expand the energy to linear order in $z$ and $\phi$:
\begin{eqnarray}
    H=\kappa _1 \phi +\gamma _1 z+E_{0}
\end{eqnarray}
We first write the coherent state with initial imbalance $z'$ and phase $\phi'$ as:
\begin{eqnarray}
    &\rho (N,z',\phi',z,\phi )=\\ \nonumber
    &\frac{\alpha _z(N,z')\alpha _{\phi }(N,z')}{\pi }e^{-\alpha _z(N,z')(z-z')^2-\alpha _{\phi }(N,z')(\phi -\phi')^2}
\end{eqnarray}
where the inverse variances are:
\begin{eqnarray}
    \alpha_{\phi}(J,z')&=& \frac{1}{2} J \left(1-z^2\right) \\ \nonumber
    \alpha_z(J,z')&=& \frac{2J}{1-z^2}
\end{eqnarray}
The eigenstate overlap is then given as:
\begin{eqnarray}
    &\rho_{diag}(z',\phi',E,s) =\\ \nonumber
    &\frac{\omega(E)}{\gamma _1}\int \text{d$\phi \rho $}\left(N,z',\phi',\frac{\delta _0-\kappa _1 \phi }{\gamma },\phi\right),
\end{eqnarray}
where $\delta_0=E-E_{0}$, and integrates to give:
\begin{eqnarray}
    \hspace{-0.8cm}\frac{\omega(E)}{\gamma _1}\frac{\sqrt{\alpha _z(N,z')\alpha _{\phi }(N,z')}}{\sqrt{\pi } \sqrt{\alpha _{\phi }+\frac{\kappa _1^2 \alpha _z}{\gamma _1^2}}}\exp \left(-\frac{\delta _0^2 \alpha _{\phi } \alpha _z}{\gamma _1^2 \alpha _{\phi }+\kappa _1^2 \alpha _z}\right)
\end{eqnarray}
Where the energy uncertainty $\sigma$ is given by:
\begin{eqnarray}
    \sigma(\phi',z') =   -\frac{\gamma _1^2 \alpha _{\phi }+\kappa _1^2 \alpha _z}{2\alpha _{\phi } \alpha _z}
\end{eqnarray}
depends on the coherent state via the uncertainties $\alpha_{z}$ and $\alpha_{\phi}$.
\section{Double Peak Saddles}
\label{apx:saddles}
Analytic solutions for the saddle point only exist if $E_{0}=1$ so we focus on coherent states on this line.
To find the saddle points we rewrite $\rho_{diag}$ as:
\begin{eqnarray}
    \rho_{diag}(E=1\pm\left|\delta\right|)=\frac{K_{\pm}}{(1 - G_{\pm}Ln[\delta])} Exp[-2JF \delta^2]
\end{eqnarray}
where $K_\pm$ and $G_{\pm}$ are constants in $\delta$, depend on $C_{\pm}$, and with $\pm$ depending on the sign of $\delta$ .
This function has a saddle at:
\begin{eqnarray}
    |\delta|=\frac{i}{\sqrt{(2JF)W_{-1}(-\frac{e^{-2/G_{\pm}}}{F 2J})}}
\end{eqnarray}
Where the product log,  $W_{-1}(X)$, is the inverse of $e^{x}x$: $W_{-1}(e^{x}x)=x$ and the $_{-1}$ says to take the negative branch.
For small $x$ we get:
\begin{eqnarray}
    \lim_{x\rightarrow0^{-}}\frac{W_{-1}(x)}{Ln(x)}=1
\end{eqnarray}
and we know $W_{-1}(x)\approx Ln(-x)-Ln(-Ln(-x))+\dots$
We therefore get the approximation:
\begin{eqnarray}
    |\delta|\approx\frac{i}{\sqrt{2JFLn(\frac{e^{-2/G_{\pm}}}{F 2J})}}
\end{eqnarray}
Which in the large-$J$ limit goes as:
\begin{eqnarray}
\frac{1}{\sqrt{2JFLn(J)}}
\end{eqnarray}
and 
\begin{eqnarray}
    2 J F=\frac{\alpha _{\phi } \alpha _z}{\gamma _1^2 \alpha _{\phi }+\kappa _1^2 \alpha _z}
\end{eqnarray}
Thus the difference in initial states on the separatrix again shows up in the scaling to the large $J$ limit.
Also note $G$ comes from $\omega(E)$ which depends on which side of the separatrix we are on (sign of $\delta$).
In the large-$J$ limit the points become symmetric as indicated by the lack of dependence on $G$.
\section{Eigenstate observables close to the separatrix}
\label{apx:obs}
Next we compute the eigenstate observables, $O(E)$, which are given as
\begin{eqnarray}
    \omega(E)\int O(z,\phi)\delta[H(z,\phi)-E].
\end{eqnarray}
$J_z$ for $\Lambda$ large has a amazingly simple solution.
For $E<1$, $J_z(E)=0$ for $E>1$ we integrate:
\begin{eqnarray}
    \int dz\delta[H(z,\phi)-E]=
    \int_{-\pi}^{\pi}d\phi z(\phi)\left|\frac{dH}{dz}\right|^{-1}
\end{eqnarray}
and for $\Lambda>>1$, $\left|\frac{dH}{dz}\right|^{-1}\approx \Lambda z$, the $z$'s cancel and we get
\begin{eqnarray}
    J_{z}(E)=\frac{\omega(E)2\pi}{\Lambda}.
\end{eqnarray}

$J_x$ is more involved.  We will take the same approach as the integral for $\omega(E)$. We assume the integral is dominated by the contribution near the unstable fixed point.
Doing so allows us to expand $J_x$ near the unstable fixed point: $J_x\approx-1+\phi^2/2$.
Solving for $\phi$, we find that it is written as: $J_x\approx \frac{\Lambda-1}{2}z^2-E$.

\begin{eqnarray}
    &J_x(1+|\delta|)^{-1}=\\ &\nonumber \omega(E)(\Lambda-1)\int_{-r_{+}}^{r_{+}} \left(-\frac{E}{\Lambda-1}+z^2/2\right)\left|\frac{dH}{dz}\right|^{-1}(\phi,\delta)+K_{+}\\\nonumber
    &J_x(1-|\delta|)^{-1}=\\ &\nonumber\omega(E)\int_{-r_{-}}^{r_{-}} (-1+\phi^2/2)\left|\frac{dH}{d\phi}\right|^{-1}(z,\delta)+K_{-}
\end{eqnarray}
Similar to the integral for $\omega(E)$, these can be computed and in the limit of small $\delta$ we get:
\begin{eqnarray}
    J_x(1+|\delta|)^{-1}&=&-1+\omega(\left|\delta\right|)\left(K_{+}-\frac{\left| \delta \right|  Ln(\left| \delta \right|)}{(\Lambda -1)^{3/2}}\right)\\\nonumber
    J_x(1-|\delta|)^{-1}&=&-1+\omega(\left|\delta\right|)\left(K_{-}-\frac{\left| \delta \right|  Ln(\left| \delta \right|)}{\sqrt{\Lambda -1}} \right)
\end{eqnarray}
$\omega(\left|\delta\right|)$ goes to $0$ faster than $\omega(\left|\delta\right|)\left|\delta\right|Ln(\left|\delta\right|)$
and we get:
\begin{eqnarray}
    J_x(1+|\delta|)^{-1}&=&-1-\omega(\left|\delta\right|)\frac{\left| \delta \right|  Ln(\left| \delta \right|)}{(\Lambda -1)^{3/2}}\\\nonumber
    J_x(1-|\delta|)^{-1}&=&-1-\omega(\left|\delta\right|)\frac{\left| \delta \right|  Ln(\left| \delta \right|)}{\sqrt{\Lambda -1}} 
\end{eqnarray}
Substituting $\omega$:
\begin{eqnarray}
    J_x(1+|\delta|)^{-1}&=&-1+ \frac{2\left| \delta \right|}{\Lambda -1}\\\nonumber
    J_x(1-|\delta|)^{-1}&=&-1+ \left| \delta \right|   \\ \nonumber
    J_{z}(1+|\delta|)&= & \frac{4\pi\sqrt{(\Lambda-1})}{\Lambda Ln\left(|\delta|\right)} \\ \nonumber
    J_{z}(1-|\delta|)&= &0
\end{eqnarray}

\bibliography{nonlin}

\begin{thebibliography}{57}%
\makeatletter
\providecommand \@ifxundefined [1]{%
 \@ifx{#1\undefined}
}%
\providecommand \@ifnum [1]{%
 \ifnum #1\expandafter \@firstoftwo
 \else \expandafter \@secondoftwo
 \fi
}%
\providecommand \@ifx [1]{%
 \ifx #1\expandafter \@firstoftwo
 \else \expandafter \@secondoftwo
 \fi
}%
\providecommand \natexlab [1]{#1}%
\providecommand \enquote  [1]{``#1''}%
\providecommand \bibnamefont  [1]{#1}%
\providecommand \bibfnamefont [1]{#1}%
\providecommand \citenamefont [1]{#1}%
\providecommand \href@noop [0]{\@secondoftwo}%
\providecommand \href [0]{\begingroup \@sanitize@url \@href}%
\providecommand \@href[1]{\@@startlink{#1}\@@href}%
\providecommand \@@href[1]{\endgroup#1\@@endlink}%
\providecommand \@sanitize@url [0]{\catcode `\\12\catcode `\$12\catcode
  `\&12\catcode `\#12\catcode `\^12\catcode `\_12\catcode `\%12\relax}%
\providecommand \@@startlink[1]{}%
\providecommand \@@endlink[0]{}%
\providecommand \url  [0]{\begingroup\@sanitize@url \@url }%
\providecommand \@url [1]{\endgroup\@href {#1}{\urlprefix }}%
\providecommand \urlprefix  [0]{URL }%
\providecommand \Eprint [0]{\href }%
\providecommand \doibase [0]{http://dx.doi.org/}%
\providecommand \selectlanguage [0]{\@gobble}%
\providecommand \bibinfo  [0]{\@secondoftwo}%
\providecommand \bibfield  [0]{\@secondoftwo}%
\providecommand \translation [1]{[#1]}%
\providecommand \BibitemOpen [0]{}%
\providecommand \bibitemStop [0]{}%
\providecommand \bibitemNoStop [0]{.\EOS\space}%
\providecommand \EOS [0]{\spacefactor3000\relax}%
\providecommand \BibitemShut  [1]{\csname bibitem#1\endcsname}%
\let\auto@bib@innerbib\@empty
%</preamble>
\bibitem [{\citenamefont {Bloch}\ \emph {et~al.}(2008)\citenamefont {Bloch},
  \citenamefont {Dalibard},\ and\ \citenamefont {Zwerger}}]{bloch2008}%
  \BibitemOpen
  \bibfield  {author} {\bibinfo {author} {\bibfnamefont {Immanuel}\
  \bibnamefont {Bloch}}, \bibinfo {author} {\bibfnamefont {Jean}\ \bibnamefont
  {Dalibard}}, \ and\ \bibinfo {author} {\bibfnamefont {Wilhelm}\ \bibnamefont
  {Zwerger}},\ }\bibfield  {title} {\enquote {\bibinfo {title} {Many-body
  physics with ultracold gases},}\ }\href {\doibase 10.1103/RevModPhys.80.885}
  {\bibfield  {journal} {\bibinfo  {journal} {Rev. Mod. Phys.}\ }\textbf
  {\bibinfo {volume} {80}},\ \bibinfo {pages} {885--964} (\bibinfo {year}
  {2008})}\BibitemShut {NoStop}%
\bibitem [{\citenamefont {Ludlow}\ \emph {et~al.}(2015)\citenamefont {Ludlow},
  \citenamefont {Boyd}, \citenamefont {Ye}, \citenamefont {Peik},\ and\
  \citenamefont {Schmidt}}]{ludlow2015}%
  \BibitemOpen
  \bibfield  {author} {\bibinfo {author} {\bibfnamefont {Andrew~D.}\
  \bibnamefont {Ludlow}}, \bibinfo {author} {\bibfnamefont {Martin~M.}\
  \bibnamefont {Boyd}}, \bibinfo {author} {\bibfnamefont {Jun}\ \bibnamefont
  {Ye}}, \bibinfo {author} {\bibfnamefont {E.}~\bibnamefont {Peik}}, \ and\
  \bibinfo {author} {\bibfnamefont {P.~O.}\ \bibnamefont {Schmidt}},\
  }\bibfield  {title} {\enquote {\bibinfo {title} {Optical atomic clocks},}\
  }\href {\doibase 10.1103/RevModPhys.87.637} {\bibfield  {journal} {\bibinfo
  {journal} {Rev. Mod. Phys.}\ }\textbf {\bibinfo {volume} {87}},\ \bibinfo
  {pages} {637--701} (\bibinfo {year} {2015})}\BibitemShut {NoStop}%
\bibitem [{\citenamefont {Shaffer}\ \emph {et~al.}(2018)\citenamefont
  {Shaffer}, \citenamefont {Rittenhouse},\ and\ \citenamefont
  {Sadeghpour}}]{shaffer2018}%
  \BibitemOpen
  \bibfield  {author} {\bibinfo {author} {\bibfnamefont {J.~P.}\ \bibnamefont
  {Shaffer}}, \bibinfo {author} {\bibfnamefont {S.~T.}\ \bibnamefont
  {Rittenhouse}}, \ and\ \bibinfo {author} {\bibfnamefont {H.~R.}\ \bibnamefont
  {Sadeghpour}},\ }\bibfield  {title} {\enquote {\bibinfo {title} {Ultracold
  {{Rydberg}} molecules},}\ }\href {\doibase 10.1038/s41467-018-04135-6}
  {\bibfield  {journal} {\bibinfo  {journal} {Nature Communications}\ }\textbf
  {\bibinfo {volume} {9}},\ \bibinfo {pages} {1965} (\bibinfo {year}
  {2018})}\BibitemShut {NoStop}%
\bibitem [{\citenamefont {Zibold}\ \emph {et~al.}(2010)\citenamefont {Zibold},
  \citenamefont {Nicklas}, \citenamefont {Gross},\ and\ \citenamefont
  {Oberthaler}}]{zibold2010}%
  \BibitemOpen
  \bibfield  {author} {\bibinfo {author} {\bibfnamefont {Tilman}\ \bibnamefont
  {Zibold}}, \bibinfo {author} {\bibfnamefont {Eike}\ \bibnamefont {Nicklas}},
  \bibinfo {author} {\bibfnamefont {Christian}\ \bibnamefont {Gross}}, \ and\
  \bibinfo {author} {\bibfnamefont {Markus~K.}\ \bibnamefont {Oberthaler}},\
  }\bibfield  {title} {\enquote {\bibinfo {title} {Classical {{Bifurcation}} at
  the {{Transition}} from {{Rabi}} to {{Josephson Dynamics}}},}\ }\href
  {\doibase 10.1103/PhysRevLett.105.204101} {\bibfield  {journal} {\bibinfo
  {journal} {Physical Review Letters}\ }\textbf {\bibinfo {volume} {105}},\
  \bibinfo {pages} {204101} (\bibinfo {year} {2010})}\BibitemShut {NoStop}%
\bibitem [{\citenamefont {Deutsch}(1991)}]{deutsch1991}%
  \BibitemOpen
  \bibfield  {author} {\bibinfo {author} {\bibfnamefont {J.~M.}\ \bibnamefont
  {Deutsch}},\ }\bibfield  {title} {\enquote {\bibinfo {title} {Quantum
  statistical mechanics in a closed system},}\ }\href {\doibase
  10.1103/PhysRevA.43.2046} {\bibfield  {journal} {\bibinfo  {journal} {Phys.
  Rev. A}\ }\textbf {\bibinfo {volume} {43}},\ \bibinfo {pages} {2046--2049}
  (\bibinfo {year} {1991})}\BibitemShut {NoStop}%
\bibitem [{\citenamefont {Srednicki}(1994)}]{srednicki1994}%
  \BibitemOpen
  \bibfield  {author} {\bibinfo {author} {\bibfnamefont {Mark}\ \bibnamefont
  {Srednicki}},\ }\bibfield  {title} {\enquote {\bibinfo {title} {Chaos and
  quantum thermalization},}\ }\href {\doibase 10.1103/PhysRevE.50.888}
  {\bibfield  {journal} {\bibinfo  {journal} {Phys. Rev. E}\ }\textbf {\bibinfo
  {volume} {50}},\ \bibinfo {pages} {888--901} (\bibinfo {year}
  {1994})}\BibitemShut {NoStop}%
\bibitem [{\citenamefont {Rigol}\ \emph {et~al.}(2008)\citenamefont {Rigol},
  \citenamefont {Dunjko},\ and\ \citenamefont {Olshanii}}]{rigol2008}%
  \BibitemOpen
  \bibfield  {author} {\bibinfo {author} {\bibfnamefont {Marcos}\ \bibnamefont
  {Rigol}}, \bibinfo {author} {\bibfnamefont {Vanja}\ \bibnamefont {Dunjko}}, \
  and\ \bibinfo {author} {\bibfnamefont {Maxim}\ \bibnamefont {Olshanii}},\
  }\bibfield  {title} {\enquote {\bibinfo {title} {Thermalization and its
  mechanism for generic isolated quantum systems},}\ }\href {\doibase
  10.1038/nature06838} {\bibfield  {journal} {\bibinfo  {journal} {Nature}\
  }\textbf {\bibinfo {volume} {452}},\ \bibinfo {pages} {854--858} (\bibinfo
  {year} {2008})}\BibitemShut {NoStop}%
\bibitem [{\citenamefont {D'Alessio}\ \emph {et~al.}(2016)\citenamefont
  {D'Alessio}, \citenamefont {Kafri}, \citenamefont {Polkovnikov},\ and\
  \citenamefont {Rigol}}]{dalessio2016a}%
  \BibitemOpen
  \bibfield  {author} {\bibinfo {author} {\bibfnamefont {Luca}\ \bibnamefont
  {D'Alessio}}, \bibinfo {author} {\bibfnamefont {Yariv}\ \bibnamefont
  {Kafri}}, \bibinfo {author} {\bibfnamefont {Anatoli}\ \bibnamefont
  {Polkovnikov}}, \ and\ \bibinfo {author} {\bibfnamefont {Marcos}\
  \bibnamefont {Rigol}},\ }\bibfield  {title} {\enquote {\bibinfo {title} {From
  quantum chaos and eigenstate thermalization to statistical mechanics and
  thermodynamics},}\ }\href {\doibase 10.1080/00018732.2016.1198134} {\bibfield
   {journal} {\bibinfo  {journal} {Advances in Physics}\ }\textbf {\bibinfo
  {volume} {65}},\ \bibinfo {pages} {239--362} (\bibinfo {year}
  {2016})}\BibitemShut {NoStop}%
\bibitem [{\citenamefont {Deutsch}(2018)}]{deutsch2018}%
  \BibitemOpen
  \bibfield  {author} {\bibinfo {author} {\bibfnamefont {Joshua~M.}\
  \bibnamefont {Deutsch}},\ }\bibfield  {title} {\enquote {\bibinfo {title}
  {Eigenstate {{Thermalization Hypothesis}}},}\ }\href {\doibase
  10.1088/1361-6633/aac9f1} {\bibfield  {journal} {\bibinfo  {journal} {Rep.
  Prog. Phys.}\ }\textbf {\bibinfo {volume} {81}},\ \bibinfo {pages} {082001}
  (\bibinfo {year} {2018})}\BibitemShut {NoStop}%
\bibitem [{\citenamefont {Jensen}\ and\ \citenamefont
  {Shankar}(1985)}]{jensen1985}%
  \BibitemOpen
  \bibfield  {author} {\bibinfo {author} {\bibfnamefont {R.~V.}\ \bibnamefont
  {Jensen}}\ and\ \bibinfo {author} {\bibfnamefont {R.}~\bibnamefont
  {Shankar}},\ }\bibfield  {title} {\enquote {\bibinfo {title} {Statistical
  {{Behavior}} in {{Deterministic Quantum Systems}} with {{Few Degrees}} of
  {{Freedom}}},}\ }\href {\doibase 10.1103/PhysRevLett.54.1879} {\bibfield
  {journal} {\bibinfo  {journal} {Phys. Rev. Lett.}\ }\textbf {\bibinfo
  {volume} {54}},\ \bibinfo {pages} {1879--1882} (\bibinfo {year}
  {1985})}\BibitemShut {NoStop}%
\bibitem [{\citenamefont {Mori}\ \emph {et~al.}(2018)\citenamefont {Mori},
  \citenamefont {Ikeda}, \citenamefont {Kaminishi},\ and\ \citenamefont
  {Ueda}}]{Mori_2018}%
  \BibitemOpen
  \bibfield  {author} {\bibinfo {author} {\bibfnamefont {Takashi}\ \bibnamefont
  {Mori}}, \bibinfo {author} {\bibfnamefont {Tatsuhiko~N}\ \bibnamefont
  {Ikeda}}, \bibinfo {author} {\bibfnamefont {Eriko}\ \bibnamefont
  {Kaminishi}}, \ and\ \bibinfo {author} {\bibfnamefont {Masahito}\
  \bibnamefont {Ueda}},\ }\bibfield  {title} {\enquote {\bibinfo {title}
  {Thermalization and prethermalization in isolated quantum systems: a
  theoretical overview},}\ }\href {\doibase 10.1088/1361-6455/aabcdf}
  {\bibfield  {journal} {\bibinfo  {journal} {Journal of Physics B: Atomic,
  Molecular and Optical Physics}\ }\textbf {\bibinfo {volume} {51}},\ \bibinfo
  {pages} {112001} (\bibinfo {year} {2018})}\BibitemShut {NoStop}%
\bibitem [{\citenamefont {Polkovnikov}\ \emph {et~al.}(2011)\citenamefont
  {Polkovnikov}, \citenamefont {Sengupta}, \citenamefont {Silva},\ and\
  \citenamefont {Vengalattore}}]{polkovnikov2011a}%
  \BibitemOpen
  \bibfield  {author} {\bibinfo {author} {\bibfnamefont {Anatoli}\ \bibnamefont
  {Polkovnikov}}, \bibinfo {author} {\bibfnamefont {Krishnendu}\ \bibnamefont
  {Sengupta}}, \bibinfo {author} {\bibfnamefont {Alessandro}\ \bibnamefont
  {Silva}}, \ and\ \bibinfo {author} {\bibfnamefont {Mukund}\ \bibnamefont
  {Vengalattore}},\ }\bibfield  {title} {\enquote {\bibinfo {title}
  {Colloquium: {{Nonequilibrium}} dynamics of closed interacting quantum
  systems},}\ }\href {\doibase 10.1103/RevModPhys.83.863} {\bibfield  {journal}
  {\bibinfo  {journal} {Rev. Mod. Phys.}\ }\textbf {\bibinfo {volume} {83}},\
  \bibinfo {pages} {863--883} (\bibinfo {year} {2011})}\BibitemShut {NoStop}%
\bibitem [{\citenamefont {Batchelor}\ and\ \citenamefont
  {Foerster}(2016)}]{batchelor2016}%
  \BibitemOpen
  \bibfield  {author} {\bibinfo {author} {\bibfnamefont {Murray~T.}\
  \bibnamefont {Batchelor}}\ and\ \bibinfo {author} {\bibfnamefont {Angela}\
  \bibnamefont {Foerster}},\ }\bibfield  {title} {\enquote {\bibinfo {title}
  {Yang\textendash{{Baxter}} integrable models in experiments: From condensed
  matter to ultracold atoms},}\ }\href {\doibase
  10.1088/1751-8113/49/17/173001} {\bibfield  {journal} {\bibinfo  {journal}
  {J. Phys. A: Math. Theor.}\ }\textbf {\bibinfo {volume} {49}},\ \bibinfo
  {pages} {173001} (\bibinfo {year} {2016})}\BibitemShut {NoStop}%
\bibitem [{\citenamefont {Cassidy}\ \emph {et~al.}(2011)\citenamefont
  {Cassidy}, \citenamefont {Clark},\ and\ \citenamefont {Rigol}}]{cassidy2011}%
  \BibitemOpen
  \bibfield  {author} {\bibinfo {author} {\bibfnamefont {Amy~C.}\ \bibnamefont
  {Cassidy}}, \bibinfo {author} {\bibfnamefont {Charles~W.}\ \bibnamefont
  {Clark}}, \ and\ \bibinfo {author} {\bibfnamefont {Marcos}\ \bibnamefont
  {Rigol}},\ }\bibfield  {title} {\enquote {\bibinfo {title} {Generalized
  {{Thermalization}} in an {{Integrable Lattice System}}},}\ }\href {\doibase
  10.1103/PhysRevLett.106.140405} {\bibfield  {journal} {\bibinfo  {journal}
  {Phys. Rev. Lett.}\ }\textbf {\bibinfo {volume} {106}},\ \bibinfo {pages}
  {140405} (\bibinfo {year} {2011})}\BibitemShut {NoStop}%
\bibitem [{\citenamefont {Imbrie}\ \emph {et~al.}(2017)\citenamefont {Imbrie},
  \citenamefont {Ros},\ and\ \citenamefont {Scardicchio}}]{imbrie2017}%
  \BibitemOpen
  \bibfield  {author} {\bibinfo {author} {\bibfnamefont {John~Z.}\ \bibnamefont
  {Imbrie}}, \bibinfo {author} {\bibfnamefont {Valentina}\ \bibnamefont {Ros}},
  \ and\ \bibinfo {author} {\bibfnamefont {Antonello}\ \bibnamefont
  {Scardicchio}},\ }\bibfield  {title} {\enquote {\bibinfo {title} {Local
  integrals of motion in many-body localized systems},}\ }\href {\doibase
  10.1002/andp.201600278} {\bibfield  {journal} {\bibinfo  {journal} {Annalen
  der Physik}\ }\textbf {\bibinfo {volume} {529}},\ \bibinfo {pages} {1600278}
  (\bibinfo {year} {2017})}\BibitemShut {NoStop}%
\bibitem [{\citenamefont {Nandkishore}\ and\ \citenamefont
  {Huse}(2015)}]{nandkishore2015}%
  \BibitemOpen
  \bibfield  {author} {\bibinfo {author} {\bibfnamefont {Rahul}\ \bibnamefont
  {Nandkishore}}\ and\ \bibinfo {author} {\bibfnamefont {David~A.}\
  \bibnamefont {Huse}},\ }\bibfield  {title} {\enquote {\bibinfo {title}
  {Many-{{Body Localization}} and {{Thermalization}} in {{Quantum Statistical
  Mechanics}}},}\ }\href {\doibase 10.1146/annurev-conmatphys-031214-014726}
  {\bibfield  {journal} {\bibinfo  {journal} {Annual Review of Condensed Matter
  Physics}\ }\textbf {\bibinfo {volume} {6}},\ \bibinfo {pages} {15--38}
  (\bibinfo {year} {2015})}\BibitemShut {NoStop}%
\bibitem [{\citenamefont {Abanin}\ \emph {et~al.}(2019)\citenamefont {Abanin},
  \citenamefont {Altman}, \citenamefont {Bloch},\ and\ \citenamefont
  {Serbyn}}]{abanin2019}%
  \BibitemOpen
  \bibfield  {author} {\bibinfo {author} {\bibfnamefont {Dmitry~A.}\
  \bibnamefont {Abanin}}, \bibinfo {author} {\bibfnamefont {Ehud}\ \bibnamefont
  {Altman}}, \bibinfo {author} {\bibfnamefont {Immanuel}\ \bibnamefont
  {Bloch}}, \ and\ \bibinfo {author} {\bibfnamefont {Maksym}\ \bibnamefont
  {Serbyn}},\ }\bibfield  {title} {\enquote {\bibinfo {title} {Colloquium:
  {{Many}}-body localization, thermalization, and entanglement},}\ }\href
  {\doibase 10.1103/RevModPhys.91.021001} {\bibfield  {journal} {\bibinfo
  {journal} {Rev. Mod. Phys.}\ }\textbf {\bibinfo {volume} {91}},\ \bibinfo
  {pages} {021001} (\bibinfo {year} {2019})}\BibitemShut {NoStop}%
\bibitem [{\citenamefont {Tikhonenkov}\ \emph {et~al.}(2013)\citenamefont
  {Tikhonenkov}, \citenamefont {Vardi}, \citenamefont {Anglin},\ and\
  \citenamefont {Cohen}}]{tikhonenkov2013}%
  \BibitemOpen
  \bibfield  {author} {\bibinfo {author} {\bibfnamefont {Igor}\ \bibnamefont
  {Tikhonenkov}}, \bibinfo {author} {\bibfnamefont {Amichay}\ \bibnamefont
  {Vardi}}, \bibinfo {author} {\bibfnamefont {James~R.}\ \bibnamefont
  {Anglin}}, \ and\ \bibinfo {author} {\bibfnamefont {Doron}\ \bibnamefont
  {Cohen}},\ }\bibfield  {title} {\enquote {\bibinfo {title} {Minimal
  {{Fokker}}-{{Planck Theory}} for the {{Thermalization}} of {{Mesoscopic
  Subsystems}}},}\ }\href {\doibase 10.1103/PhysRevLett.110.050401} {\bibfield
  {journal} {\bibinfo  {journal} {Phys. Rev. Lett.}\ }\textbf {\bibinfo
  {volume} {110}},\ \bibinfo {pages} {050401} (\bibinfo {year}
  {2013})}\BibitemShut {NoStop}%
\bibitem [{\citenamefont {Arwas}\ \emph {et~al.}(2015)\citenamefont {Arwas},
  \citenamefont {Vardi},\ and\ \citenamefont {Cohen}}]{arwas2015a}%
  \BibitemOpen
  \bibfield  {author} {\bibinfo {author} {\bibfnamefont {Geva}\ \bibnamefont
  {Arwas}}, \bibinfo {author} {\bibfnamefont {Amichay}\ \bibnamefont {Vardi}},
  \ and\ \bibinfo {author} {\bibfnamefont {Doron}\ \bibnamefont {Cohen}},\
  }\bibfield  {title} {\enquote {\bibinfo {title} {Superfluidity and {{Chaos}}
  in low dimensional circuits},}\ }\href {\doibase 10.1038/srep13433}
  {\bibfield  {journal} {\bibinfo  {journal} {Scientific Reports}\ }\textbf
  {\bibinfo {volume} {5}},\ \bibinfo {pages} {13433} (\bibinfo {year}
  {2015})}\BibitemShut {NoStop}%
\bibitem [{\citenamefont {Khripkov}\ \emph {et~al.}(2016)\citenamefont
  {Khripkov}, \citenamefont {Cohen},\ and\ \citenamefont
  {Vardi}}]{khripkov2016a}%
  \BibitemOpen
  \bibfield  {author} {\bibinfo {author} {\bibfnamefont {Christine}\
  \bibnamefont {Khripkov}}, \bibinfo {author} {\bibfnamefont {Doron}\
  \bibnamefont {Cohen}}, \ and\ \bibinfo {author} {\bibfnamefont {Amichay}\
  \bibnamefont {Vardi}},\ }\bibfield  {title} {\enquote {\bibinfo {title}
  {Thermalization of {{Bipartite Bose}}\textendash{{Hubbard Models}}},}\ }\href
  {\doibase 10.1021/acs.jpca.5b11176} {\bibfield  {journal} {\bibinfo
  {journal} {J. Phys. Chem. A}\ }\textbf {\bibinfo {volume} {120}},\ \bibinfo
  {pages} {3136--3141} (\bibinfo {year} {2016})}\BibitemShut {NoStop}%
\bibitem [{\citenamefont {Khripkov}\ \emph {et~al.}(2018)\citenamefont
  {Khripkov}, \citenamefont {Vardi},\ and\ \citenamefont
  {Cohen}}]{khripkov2018a}%
  \BibitemOpen
  \bibfield  {author} {\bibinfo {author} {\bibfnamefont {Christine}\
  \bibnamefont {Khripkov}}, \bibinfo {author} {\bibfnamefont {Amichay}\
  \bibnamefont {Vardi}}, \ and\ \bibinfo {author} {\bibfnamefont {Doron}\
  \bibnamefont {Cohen}},\ }\bibfield  {title} {\enquote {\bibinfo {title}
  {Semiclassical theory of strong localization for quantum thermalization},}\
  }\href {\doibase 10.1103/PhysRevE.97.022127} {\bibfield  {journal} {\bibinfo
  {journal} {Phys. Rev. E}\ }\textbf {\bibinfo {volume} {97}},\ \bibinfo
  {pages} {022127} (\bibinfo {year} {2018})}\BibitemShut {NoStop}%
\bibitem [{\citenamefont {Arwas}\ and\ \citenamefont
  {Cohen}(2016)}]{arwas2016}%
  \BibitemOpen
  \bibfield  {author} {\bibinfo {author} {\bibfnamefont {Geva}\ \bibnamefont
  {Arwas}}\ and\ \bibinfo {author} {\bibfnamefont {Doron}\ \bibnamefont
  {Cohen}},\ }\bibfield  {title} {\enquote {\bibinfo {title} {Chaos and
  two-level dynamics of the atomtronic quantum interference device},}\ }\href
  {\doibase 10.1088/1367-2630/18/1/015007} {\bibfield  {journal} {\bibinfo
  {journal} {New J. Phys.}\ }\textbf {\bibinfo {volume} {18}},\ \bibinfo
  {pages} {015007} (\bibinfo {year} {2016})}\BibitemShut {NoStop}%
\bibitem [{\citenamefont {Arwas}\ and\ \citenamefont
  {Cohen}(2017)}]{arwas2017a}%
  \BibitemOpen
  \bibfield  {author} {\bibinfo {author} {\bibfnamefont {Geva}\ \bibnamefont
  {Arwas}}\ and\ \bibinfo {author} {\bibfnamefont {Doron}\ \bibnamefont
  {Cohen}},\ }\bibfield  {title} {\enquote {\bibinfo {title} {Chaos,
  metastability and ergodicity in {{Bose}}-{{Hubbard}} superfluid circuits},}\
  }\href {\doibase 10.1063/1.5016126} {\bibfield  {journal} {\bibinfo
  {journal} {AIP Conference Proceedings}\ }\textbf {\bibinfo {volume} {1912}},\
  \bibinfo {pages} {020001} (\bibinfo {year} {2017})}\BibitemShut {NoStop}%
\bibitem [{\citenamefont {Arwas}\ and\ \citenamefont
  {Cohen}(2019)}]{arwas2019a}%
  \BibitemOpen
  \bibfield  {author} {\bibinfo {author} {\bibfnamefont {Geva}\ \bibnamefont
  {Arwas}}\ and\ \bibinfo {author} {\bibfnamefont {Doron}\ \bibnamefont
  {Cohen}},\ }\bibfield  {title} {\enquote {\bibinfo {title} {Monodromy and
  chaos for condensed bosons in optical lattices},}\ }\href {\doibase
  10.1103/PhysRevA.99.023625} {\bibfield  {journal} {\bibinfo  {journal} {Phys.
  Rev. A}\ }\textbf {\bibinfo {volume} {99}},\ \bibinfo {pages} {023625}
  (\bibinfo {year} {2019})}\BibitemShut {NoStop}%
\bibitem [{\citenamefont {Pizzi}\ \emph
  {et~al.}(2019{\natexlab{a}})\citenamefont {Pizzi}, \citenamefont {Dolcini},\
  and\ \citenamefont {Le~Hur}}]{PiziiFixPOints}%
  \BibitemOpen
  \bibfield  {author} {\bibinfo {author} {\bibfnamefont {Andrea}\ \bibnamefont
  {Pizzi}}, \bibinfo {author} {\bibfnamefont {Fabrizio}\ \bibnamefont
  {Dolcini}}, \ and\ \bibinfo {author} {\bibfnamefont {Karyn}\ \bibnamefont
  {Le~Hur}},\ }\bibfield  {title} {\enquote {\bibinfo {title} {Quench-induced
  dynamical phase transitions and $\ensuremath{\pi}$-synchronization in the
  bose-hubbard model},}\ }\href {\doibase 10.1103/PhysRevB.99.094301}
  {\bibfield  {journal} {\bibinfo  {journal} {Phys. Rev. B}\ }\textbf {\bibinfo
  {volume} {99}},\ \bibinfo {pages} {094301} (\bibinfo {year}
  {2019}{\natexlab{a}})}\BibitemShut {NoStop}%
\bibitem [{\citenamefont {Pizzi}\ \emph
  {et~al.}(2019{\natexlab{b}})\citenamefont {Pizzi}, \citenamefont {Knolle},\
  and\ \citenamefont {Nunnenkamp}}]{PizziTimeCrystal}%
  \BibitemOpen
  \bibfield  {author} {\bibinfo {author} {\bibfnamefont {Andrea}\ \bibnamefont
  {Pizzi}}, \bibinfo {author} {\bibfnamefont {Johannes}\ \bibnamefont
  {Knolle}}, \ and\ \bibinfo {author} {\bibfnamefont {Andreas}\ \bibnamefont
  {Nunnenkamp}},\ }\bibfield  {title} {\enquote {\bibinfo {title} {Period-$n$
  discrete time crystals and quasicrystals with ultracold bosons},}\ }\href
  {\doibase 10.1103/PhysRevLett.123.150601} {\bibfield  {journal} {\bibinfo
  {journal} {Phys. Rev. Lett.}\ }\textbf {\bibinfo {volume} {123}},\ \bibinfo
  {pages} {150601} (\bibinfo {year} {2019}{\natexlab{b}})}\BibitemShut
  {NoStop}%
\bibitem [{\citenamefont {Strobel}\ \emph {et~al.}(2014)\citenamefont
  {Strobel}, \citenamefont {Muessel}, \citenamefont {Linnemann}, \citenamefont
  {Zibold}, \citenamefont {Hume}, \citenamefont {Pezze'}, \citenamefont
  {Smerzi},\ and\ \citenamefont {Oberthaler}}]{strobel2014}%
  \BibitemOpen
  \bibfield  {author} {\bibinfo {author} {\bibfnamefont {Helmut}\ \bibnamefont
  {Strobel}}, \bibinfo {author} {\bibfnamefont {Wolfgang}\ \bibnamefont
  {Muessel}}, \bibinfo {author} {\bibfnamefont {Daniel}\ \bibnamefont
  {Linnemann}}, \bibinfo {author} {\bibfnamefont {Tilman}\ \bibnamefont
  {Zibold}}, \bibinfo {author} {\bibfnamefont {David~B.}\ \bibnamefont {Hume}},
  \bibinfo {author} {\bibfnamefont {Luca}\ \bibnamefont {Pezze'}}, \bibinfo
  {author} {\bibfnamefont {Augusto}\ \bibnamefont {Smerzi}}, \ and\ \bibinfo
  {author} {\bibfnamefont {Markus~K.}\ \bibnamefont {Oberthaler}},\ }\bibfield
  {title} {\enquote {\bibinfo {title} {Fisher information and entanglement of
  non-{{Gaussian}} spin states},}\ }\href {\doibase 10.1126/science.1250147}
  {\bibfield  {journal} {\bibinfo  {journal} {Science}\ }\textbf {\bibinfo
  {volume} {345}},\ \bibinfo {pages} {424--427} (\bibinfo {year}
  {2014})}\BibitemShut {NoStop}%
\bibitem [{\citenamefont {Raghavan}\ \emph {et~al.}(1999)\citenamefont
  {Raghavan}, \citenamefont {Smerzi}, \citenamefont {Fantoni},\ and\
  \citenamefont {Shenoy}}]{raghavan1999}%
  \BibitemOpen
  \bibfield  {author} {\bibinfo {author} {\bibfnamefont {S.}~\bibnamefont
  {Raghavan}}, \bibinfo {author} {\bibfnamefont {A.}~\bibnamefont {Smerzi}},
  \bibinfo {author} {\bibfnamefont {S.}~\bibnamefont {Fantoni}}, \ and\
  \bibinfo {author} {\bibfnamefont {S.~R.}\ \bibnamefont {Shenoy}},\ }\bibfield
   {title} {\enquote {\bibinfo {title} {Coherent oscillations between two
  weakly coupled {{Bose}}-{{Einstein}} condensates: {{Josephson}} effects, pi
  oscillations, and macroscopic quantum self-trapping},}\ }\href {\doibase
  10.1103/PhysRevA.59.620} {\bibfield  {journal} {\bibinfo  {journal} {Physical
  Review A}\ }\textbf {\bibinfo {volume} {59}},\ \bibinfo {pages} {620--633}
  (\bibinfo {year} {1999})}\BibitemShut {NoStop}%
\bibitem [{\citenamefont {Micheli}\ \emph {et~al.}(2003)\citenamefont
  {Micheli}, \citenamefont {Jaksch}, \citenamefont {Cirac},\ and\ \citenamefont
  {Zoller}}]{micheli2003}%
  \BibitemOpen
  \bibfield  {author} {\bibinfo {author} {\bibfnamefont {A.}~\bibnamefont
  {Micheli}}, \bibinfo {author} {\bibfnamefont {D.}~\bibnamefont {Jaksch}},
  \bibinfo {author} {\bibfnamefont {J.~I.}\ \bibnamefont {Cirac}}, \ and\
  \bibinfo {author} {\bibfnamefont {P.}~\bibnamefont {Zoller}},\ }\bibfield
  {title} {\enquote {\bibinfo {title} {Many-particle entanglement in
  two-component {{Bose}}-{{Einstein}} condensates},}\ }\href {\doibase
  10.1103/PhysRevA.67.013607} {\bibfield  {journal} {\bibinfo  {journal}
  {Physical Review A}\ }\textbf {\bibinfo {volume} {67}},\ \bibinfo {pages}
  {013607} (\bibinfo {year} {2003})}\BibitemShut {NoStop}%
\bibitem [{\citenamefont {Mahmud}\ \emph {et~al.}(2005)\citenamefont {Mahmud},
  \citenamefont {Perry},\ and\ \citenamefont {Reinhardt}}]{mahmud2005}%
  \BibitemOpen
  \bibfield  {author} {\bibinfo {author} {\bibfnamefont {Khan~W.}\ \bibnamefont
  {Mahmud}}, \bibinfo {author} {\bibfnamefont {Heidi}\ \bibnamefont {Perry}}, \
  and\ \bibinfo {author} {\bibfnamefont {William~P.}\ \bibnamefont
  {Reinhardt}},\ }\bibfield  {title} {\enquote {\bibinfo {title} {Quantum
  phase-space picture of {{Bose}}-{{Einstein}} condensates in a double well},}\
  }\href {\doibase 10.1103/PhysRevA.71.023615} {\bibfield  {journal} {\bibinfo
  {journal} {Physical Review A}\ }\textbf {\bibinfo {volume} {71}},\ \bibinfo
  {pages} {023615} (\bibinfo {year} {2005})}\BibitemShut {NoStop}%
\bibitem [{\citenamefont {Chuchem}\ \emph {et~al.}(2010)\citenamefont
  {Chuchem}, \citenamefont {{Smith-Mannschott}}, \citenamefont {Hiller},
  \citenamefont {Kottos}, \citenamefont {Vardi},\ and\ \citenamefont
  {Cohen}}]{chuchem2010}%
  \BibitemOpen
  \bibfield  {author} {\bibinfo {author} {\bibfnamefont {Maya}\ \bibnamefont
  {Chuchem}}, \bibinfo {author} {\bibfnamefont {Katrina}\ \bibnamefont
  {{Smith-Mannschott}}}, \bibinfo {author} {\bibfnamefont {Moritz}\
  \bibnamefont {Hiller}}, \bibinfo {author} {\bibfnamefont {Tsampikos}\
  \bibnamefont {Kottos}}, \bibinfo {author} {\bibfnamefont {Amichay}\
  \bibnamefont {Vardi}}, \ and\ \bibinfo {author} {\bibfnamefont {Doron}\
  \bibnamefont {Cohen}},\ }\bibfield  {title} {\enquote {\bibinfo {title}
  {Quantum dynamics in the bosonic {{Josephson}} junction},}\ }\href {\doibase
  10.1103/PhysRevA.82.053617} {\bibfield  {journal} {\bibinfo  {journal}
  {Physical Review A}\ }\textbf {\bibinfo {volume} {82}},\ \bibinfo {pages}
  {053617} (\bibinfo {year} {2010})}\BibitemShut {NoStop}%
\bibitem [{\citenamefont {Huang}\ \emph {et~al.}(2012)\citenamefont {Huang},
  \citenamefont {Zhong}, \citenamefont {Sun},\ and\ \citenamefont
  {Wang}}]{huang2012}%
  \BibitemOpen
  \bibfield  {author} {\bibinfo {author} {\bibfnamefont {Yixiao}\ \bibnamefont
  {Huang}}, \bibinfo {author} {\bibfnamefont {Wei}\ \bibnamefont {Zhong}},
  \bibinfo {author} {\bibfnamefont {Zhe}\ \bibnamefont {Sun}}, \ and\ \bibinfo
  {author} {\bibfnamefont {Xiaoguang}\ \bibnamefont {Wang}},\ }\bibfield
  {title} {\enquote {\bibinfo {title} {Fisher-information manifestation of
  dynamical stability and transition to self-trapping for {{Bose}}-{{Einstein}}
  condensates},}\ }\href {\doibase 10.1103/PhysRevA.86.012320} {\bibfield
  {journal} {\bibinfo  {journal} {Physical Review A - Atomic, Molecular, and
  Optical Physics}\ }\textbf {\bibinfo {volume} {86}},\ \bibinfo {pages} {1--7}
  (\bibinfo {year} {2012})}\BibitemShut {NoStop}%
\bibitem [{\citenamefont {Lapert}\ \emph {et~al.}(2012)\citenamefont {Lapert},
  \citenamefont {Ferrini},\ and\ \citenamefont {Sugny}}]{lapert2012}%
  \BibitemOpen
  \bibfield  {author} {\bibinfo {author} {\bibfnamefont {M.}~\bibnamefont
  {Lapert}}, \bibinfo {author} {\bibfnamefont {G.}~\bibnamefont {Ferrini}}, \
  and\ \bibinfo {author} {\bibfnamefont {D.}~\bibnamefont {Sugny}},\ }\bibfield
   {title} {\enquote {\bibinfo {title} {Optimal control of quantum
  superpositions in a bosonic {{Josephson}} junction},}\ }\href {\doibase
  10.1103/PhysRevA.85.023611} {\bibfield  {journal} {\bibinfo  {journal}
  {Physical Review A - Atomic, Molecular, and Optical Physics}\ }\textbf
  {\bibinfo {volume} {85}},\ \bibinfo {pages} {1--13} (\bibinfo {year}
  {2012})}\BibitemShut {NoStop}%
\bibitem [{\citenamefont {Khripkov}\ \emph {et~al.}(2013)\citenamefont
  {Khripkov}, \citenamefont {Cohen},\ and\ \citenamefont
  {Vardi}}]{khripkov2013}%
  \BibitemOpen
  \bibfield  {author} {\bibinfo {author} {\bibfnamefont {Christine}\
  \bibnamefont {Khripkov}}, \bibinfo {author} {\bibfnamefont {Doron}\
  \bibnamefont {Cohen}}, \ and\ \bibinfo {author} {\bibfnamefont {Amichay}\
  \bibnamefont {Vardi}},\ }\bibfield  {title} {\enquote {\bibinfo {title}
  {Temporal fluctuations in the bosonic {{Josephson}} junction as a probe for
  phase space tomography},}\ }\href {\doibase 10.1088/1751-8113/46/16/165304}
  {\bibfield  {journal} {\bibinfo  {journal} {J. Phys. A: Math. Theor.}\
  }\textbf {\bibinfo {volume} {46}},\ \bibinfo {pages} {165304} (\bibinfo
  {year} {2013})}\BibitemShut {NoStop}%
\bibitem [{\citenamefont {Lovas}\ \emph {et~al.}(2017)\citenamefont {Lovas},
  \citenamefont {Fort{\'a}gh}, \citenamefont {Demler},\ and\ \citenamefont
  {Zar{\'a}nd}}]{lovas2017}%
  \BibitemOpen
  \bibfield  {author} {\bibinfo {author} {\bibfnamefont {Izabella}\
  \bibnamefont {Lovas}}, \bibinfo {author} {\bibfnamefont {J{\'o}zsef}\
  \bibnamefont {Fort{\'a}gh}}, \bibinfo {author} {\bibfnamefont {Eugene}\
  \bibnamefont {Demler}}, \ and\ \bibinfo {author} {\bibfnamefont {Gergely}\
  \bibnamefont {Zar{\'a}nd}},\ }\bibfield  {title} {\enquote {\bibinfo {title}
  {Entanglement and entropy production in coupled single-mode
  {{Bose}}-{{Einstein}} condensates},}\ }\href {\doibase
  10.1103/PhysRevA.96.023615} {\bibfield  {journal} {\bibinfo  {journal} {Phys.
  Rev. A}\ }\textbf {\bibinfo {volume} {96}},\ \bibinfo {pages} {023615}
  (\bibinfo {year} {2017})}\BibitemShut {NoStop}%
\bibitem [{\citenamefont {Mathew}\ and\ \citenamefont
  {Tiesinga}(2017)}]{mathew2017}%
  \BibitemOpen
  \bibfield  {author} {\bibinfo {author} {\bibfnamefont {R.}~\bibnamefont
  {Mathew}}\ and\ \bibinfo {author} {\bibfnamefont {E.}~\bibnamefont
  {Tiesinga}},\ }\bibfield  {title} {\enquote {\bibinfo {title} {Phase-space
  mixing in dynamically unstable, integrable few-mode quantum systems},}\
  }\href {\doibase 10.1103/PhysRevA.96.013604} {\bibfield  {journal} {\bibinfo
  {journal} {Phys. Rev. A}\ }\textbf {\bibinfo {volume} {96}},\ \bibinfo
  {pages} {013604} (\bibinfo {year} {2017})}\BibitemShut {NoStop}%
\bibitem [{\citenamefont {Kelly}\ \emph {et~al.}(2019)\citenamefont {Kelly},
  \citenamefont {Timmermans},\ and\ \citenamefont {Tsai}}]{kelly2019}%
  \BibitemOpen
  \bibfield  {author} {\bibinfo {author} {\bibfnamefont {Shane~P.}\
  \bibnamefont {Kelly}}, \bibinfo {author} {\bibfnamefont {Eddy}\ \bibnamefont
  {Timmermans}}, \ and\ \bibinfo {author} {\bibfnamefont {S.-W.}\ \bibnamefont
  {Tsai}},\ }\bibfield  {title} {\enquote {\bibinfo {title} {Detecting
  macroscopic indefiniteness of cat states in bosonic interferometers},}\
  }\href {\doibase 10.1103/PhysRevA.100.032117} {\bibfield  {journal} {\bibinfo
   {journal} {Phys. Rev. A}\ }\textbf {\bibinfo {volume} {100}},\ \bibinfo
  {pages} {032117} (\bibinfo {year} {2019})}\BibitemShut {NoStop}%
\bibitem [{\citenamefont {Hennig}\ \emph {et~al.}(2012)\citenamefont {Hennig},
  \citenamefont {Witthaut},\ and\ \citenamefont {Campbell}}]{hennig20212}%
  \BibitemOpen
  \bibfield  {author} {\bibinfo {author} {\bibfnamefont {Holger}\ \bibnamefont
  {Hennig}}, \bibinfo {author} {\bibfnamefont {Dirk}\ \bibnamefont {Witthaut}},
  \ and\ \bibinfo {author} {\bibfnamefont {David~K.}\ \bibnamefont
  {Campbell}},\ }\bibfield  {title} {\enquote {\bibinfo {title} {Global phase
  space of coherence and entanglement in a double-well {{Bose}}-{{Einstein}}
  condensate},}\ }\href {\doibase 10.1103/PhysRevA.86.051604} {\bibfield
  {journal} {\bibinfo  {journal} {Phys. Rev. A}\ }\textbf {\bibinfo {volume}
  {86}},\ \bibinfo {pages} {051604} (\bibinfo {year} {2012})}\BibitemShut
  {NoStop}%
\bibitem [{\citenamefont {Morita}\ \emph {et~al.}(2006)\citenamefont {Morita},
  \citenamefont {Ohnishi}, \citenamefont {[da Provid{\^e}ncia]},\ and\
  \citenamefont {Nishiyama}}]{MORITA2006337}%
  \BibitemOpen
  \bibfield  {author} {\bibinfo {author} {\bibfnamefont {Hiroyuki}\
  \bibnamefont {Morita}}, \bibinfo {author} {\bibfnamefont {Hiromasa}\
  \bibnamefont {Ohnishi}}, \bibinfo {author} {\bibfnamefont {Jo{\~a}o}\
  \bibnamefont {[da Provid{\^e}ncia]}}, \ and\ \bibinfo {author} {\bibfnamefont
  {Seiya}\ \bibnamefont {Nishiyama}},\ }\bibfield  {title} {\enquote {\bibinfo
  {title} {Exact solutions for the {LMG} model {Hamiltonian} based on the
  {Bethe} ansatz},}\ }\href {\doibase
  https://doi.org/10.1016/j.nuclphysb.2006.01.015} {\bibfield  {journal}
  {\bibinfo  {journal} {Nuclear Physics B}\ }\textbf {\bibinfo {volume}
  {737}},\ \bibinfo {pages} {337 -- 350} (\bibinfo {year} {2006})}\BibitemShut
  {NoStop}%
\bibitem [{\citenamefont {Polkovnikov}(2010)}]{polkovnikov2009}%
  \BibitemOpen
  \bibfield  {author} {\bibinfo {author} {\bibfnamefont {Anatoli}\ \bibnamefont
  {Polkovnikov}},\ }\bibfield  {title} {\enquote {\bibinfo {title} {Phase space
  representation of quantum dynamics},}\ }\href {\doibase
  https://doi.org/10.1016/j.aop.2010.02.006} {\bibfield  {journal} {\bibinfo
  {journal} {Annals of Physics}\ }\textbf {\bibinfo {volume} {325}},\ \bibinfo
  {pages} {1790 -- 1852} (\bibinfo {year} {2010})}\BibitemShut {NoStop}%
\bibitem [{\citenamefont {Mori}(2017)}]{Mori2017}%
  \BibitemOpen
  \bibfield  {author} {\bibinfo {author} {\bibfnamefont {Takashi}\ \bibnamefont
  {Mori}},\ }\bibfield  {title} {\enquote {\bibinfo {title} {Classical
  ergodicity and quantum eigenstate thermalization: Analysis in fully connected
  ising ferromagnets},}\ }\href {\doibase 10.1103/PhysRevE.96.012134}
  {\bibfield  {journal} {\bibinfo  {journal} {Phys. Rev. E}\ }\textbf {\bibinfo
  {volume} {96}},\ \bibinfo {pages} {012134} (\bibinfo {year}
  {2017})}\BibitemShut {NoStop}%
\bibitem [{\citenamefont {Albiez}\ \emph {et~al.}(2005)\citenamefont {Albiez},
  \citenamefont {Gati}, \citenamefont {F{\"o}lling}, \citenamefont {Hunsmann},
  \citenamefont {Cristiani},\ and\ \citenamefont {Oberthaler}}]{albiez2005}%
  \BibitemOpen
  \bibfield  {author} {\bibinfo {author} {\bibfnamefont {Michael}\ \bibnamefont
  {Albiez}}, \bibinfo {author} {\bibfnamefont {Rudolf}\ \bibnamefont {Gati}},
  \bibinfo {author} {\bibfnamefont {Jonas}\ \bibnamefont {F{\"o}lling}},
  \bibinfo {author} {\bibfnamefont {Stefan}\ \bibnamefont {Hunsmann}}, \bibinfo
  {author} {\bibfnamefont {Matteo}\ \bibnamefont {Cristiani}}, \ and\ \bibinfo
  {author} {\bibfnamefont {Markus~K.}\ \bibnamefont {Oberthaler}},\ }\bibfield
  {title} {\enquote {\bibinfo {title} {Direct observation of tunneling and
  nonlinear self-trapping in a single bosonic josephson junction},}\ }\href
  {\doibase 10.1103/PhysRevLett.95.010402} {\bibfield  {journal} {\bibinfo
  {journal} {Physical Review Letters}\ }\textbf {\bibinfo {volume} {95}},\
  \bibinfo {pages} {010402--010402} (\bibinfo {year} {2005})}\BibitemShut
  {NoStop}%
\bibitem [{\citenamefont {Sakurai}(1994)}]{sakurai1994modern}%
  \BibitemOpen
  \bibfield  {author} {\bibinfo {author} {\bibfnamefont {J.~J.}\ \bibnamefont
  {Sakurai}},\ }\href@noop {} {\emph {\bibinfo {title} {Modern quantum
  mechanics}}}\ (\bibinfo  {publisher} {Addison-Wesley Pub. Co},\ \bibinfo
  {address} {Reading, Mass},\ \bibinfo {year} {1994})\BibitemShut {NoStop}%
\bibitem [{\citenamefont {Pudlik}\ \emph {et~al.}(2014)\citenamefont {Pudlik},
  \citenamefont {Hennig}, \citenamefont {Witthaut},\ and\ \citenamefont
  {Campbell}}]{pudlik2014}%
  \BibitemOpen
  \bibfield  {author} {\bibinfo {author} {\bibfnamefont {Tadeusz}\ \bibnamefont
  {Pudlik}}, \bibinfo {author} {\bibfnamefont {Holger}\ \bibnamefont {Hennig}},
  \bibinfo {author} {\bibfnamefont {Dirk}\ \bibnamefont {Witthaut}}, \ and\
  \bibinfo {author} {\bibfnamefont {David~K.}\ \bibnamefont {Campbell}},\
  }\bibfield  {title} {\enquote {\bibinfo {title} {Tunneling in the
  self-trapped regime of a two-well {{Bose}}-{{Einstein}} condensate},}\ }\href
  {\doibase 10.1103/PhysRevA.90.053610} {\bibfield  {journal} {\bibinfo
  {journal} {Phys. Rev. A}\ }\textbf {\bibinfo {volume} {90}},\ \bibinfo
  {pages} {053610} (\bibinfo {year} {2014})}\BibitemShut {NoStop}%
\bibitem [{\citenamefont {Benitez}\ \emph {et~al.}(2009)\citenamefont
  {Benitez}, \citenamefont {{Romero-Rochin}},\ and\ \citenamefont
  {Paredes}}]{benitez2009}%
  \BibitemOpen
  \bibfield  {author} {\bibinfo {author} {\bibfnamefont {S.~F.~Caballero}\
  \bibnamefont {Benitez}}, \bibinfo {author} {\bibfnamefont {V.}~\bibnamefont
  {{Romero-Rochin}}}, \ and\ \bibinfo {author} {\bibfnamefont {R.}~\bibnamefont
  {Paredes}},\ }\bibfield  {title} {\enquote {\bibinfo {title} {Delocalization
  to self-trapping transition of a {{Bose}} fluid confined in a double well
  potential. {{An}} analysis via one- and two-body correlation properties},}\
  }\href {\doibase 10.1088/0953-4075/43/11/115301} {\bibfield  {journal}
  {\bibinfo  {journal} {Journal of Physics B: Atomic, Molecular and Optical
  Physics}\ }\textbf {\bibinfo {volume} {43}},\ \bibinfo {pages}
  {115301--115301} (\bibinfo {year} {2009})}\BibitemShut {NoStop}%
\bibitem [{\citenamefont {Pilatowsky-Cameo}\ \emph {et~al.}(2020)\citenamefont
  {Pilatowsky-Cameo}, \citenamefont {Ch\'avez-Carlos}, \citenamefont
  {Bastarrachea-Magnani}, \citenamefont {Str\'ansk\'y}, \citenamefont
  {Lerma-Hern\'andez}, \citenamefont {Santos},\ and\ \citenamefont
  {Hirsch}}]{Pilatowsky2020}%
  \BibitemOpen
  \bibfield  {author} {\bibinfo {author} {\bibfnamefont {Sa\'ul}\ \bibnamefont
  {Pilatowsky-Cameo}}, \bibinfo {author} {\bibfnamefont {Jorge}\ \bibnamefont
  {Ch\'avez-Carlos}}, \bibinfo {author} {\bibfnamefont {Miguel~A.}\
  \bibnamefont {Bastarrachea-Magnani}}, \bibinfo {author} {\bibfnamefont
  {Pavel}\ \bibnamefont {Str\'ansk\'y}}, \bibinfo {author} {\bibfnamefont
  {Sergio}\ \bibnamefont {Lerma-Hern\'andez}}, \bibinfo {author} {\bibfnamefont
  {Lea~F.}\ \bibnamefont {Santos}}, \ and\ \bibinfo {author} {\bibfnamefont
  {Jorge~G.}\ \bibnamefont {Hirsch}},\ }\bibfield  {title} {\enquote {\bibinfo
  {title} {Positive quantum lyapunov exponents in experimental systems with a
  regular classical limit},}\ }\href {\doibase 10.1103/PhysRevE.101.010202}
  {\bibfield  {journal} {\bibinfo  {journal} {Phys. Rev. E}\ }\textbf {\bibinfo
  {volume} {101}},\ \bibinfo {pages} {010202} (\bibinfo {year}
  {2020})}\BibitemShut {NoStop}%
\bibitem [{\citenamefont {Xu}\ \emph {et~al.}(2020)\citenamefont {Xu},
  \citenamefont {Scaffidi},\ and\ \citenamefont {Cao}}]{Tianrui2020}%
  \BibitemOpen
  \bibfield  {author} {\bibinfo {author} {\bibfnamefont {Tianrui}\ \bibnamefont
  {Xu}}, \bibinfo {author} {\bibfnamefont {Thomas}\ \bibnamefont {Scaffidi}}, \
  and\ \bibinfo {author} {\bibfnamefont {Xiangyu}\ \bibnamefont {Cao}},\
  }\bibfield  {title} {\enquote {\bibinfo {title} {Does scrambling equal
  chaos?}}\ }\href {\doibase 10.1103/PhysRevLett.124.140602} {\bibfield
  {journal} {\bibinfo  {journal} {Phys. Rev. Lett.}\ }\textbf {\bibinfo
  {volume} {124}},\ \bibinfo {pages} {140602} (\bibinfo {year}
  {2020})}\BibitemShut {NoStop}%
\bibitem [{\citenamefont {Rozenbaum}\ \emph {et~al.}(2020)\citenamefont
  {Rozenbaum}, \citenamefont {Bunimovich},\ and\ \citenamefont
  {Galitski}}]{Rozenbaum2020}%
  \BibitemOpen
  \bibfield  {author} {\bibinfo {author} {\bibfnamefont {Efim~B.}\ \bibnamefont
  {Rozenbaum}}, \bibinfo {author} {\bibfnamefont {Leonid~A.}\ \bibnamefont
  {Bunimovich}}, \ and\ \bibinfo {author} {\bibfnamefont {Victor}\ \bibnamefont
  {Galitski}},\ }\bibfield  {title} {\enquote {\bibinfo {title} {Early-time
  exponential instabilities in nonchaotic quantum systems},}\ }\href {\doibase
  10.1103/PhysRevLett.125.014101} {\bibfield  {journal} {\bibinfo  {journal}
  {Phys. Rev. Lett.}\ }\textbf {\bibinfo {volume} {125}},\ \bibinfo {pages}
  {014101} (\bibinfo {year} {2020})}\BibitemShut {NoStop}%
\bibitem [{\citenamefont {Lashkari}\ \emph {et~al.}(2013)\citenamefont
  {Lashkari}, \citenamefont {Stanford}, \citenamefont {Hastings}, \citenamefont
  {Osborne},\ and\ \citenamefont {Hayden}}]{lashkari2013}%
  \BibitemOpen
  \bibfield  {author} {\bibinfo {author} {\bibfnamefont {Nima}\ \bibnamefont
  {Lashkari}}, \bibinfo {author} {\bibfnamefont {Douglas}\ \bibnamefont
  {Stanford}}, \bibinfo {author} {\bibfnamefont {Matthew}\ \bibnamefont
  {Hastings}}, \bibinfo {author} {\bibfnamefont {Tobias}\ \bibnamefont
  {Osborne}}, \ and\ \bibinfo {author} {\bibfnamefont {Patrick}\ \bibnamefont
  {Hayden}},\ }\bibfield  {title} {\enquote {\bibinfo {title} {Towards the fast
  scrambling conjecture},}\ }\href {\doibase 10.1007/JHEP04(2013)022}
  {\bibfield  {journal} {\bibinfo  {journal} {J. High Energ. Phys.}\ }\textbf
  {\bibinfo {volume} {2013}},\ \bibinfo {pages} {22} (\bibinfo {year}
  {2013})}\BibitemShut {NoStop}%
\bibitem [{\citenamefont {Maldacena}\ \emph {et~al.}(2016)\citenamefont
  {Maldacena}, \citenamefont {Shenker},\ and\ \citenamefont
  {Stanford}}]{maldacena2016a}%
  \BibitemOpen
  \bibfield  {author} {\bibinfo {author} {\bibfnamefont {Juan}\ \bibnamefont
  {Maldacena}}, \bibinfo {author} {\bibfnamefont {Stephen~H.}\ \bibnamefont
  {Shenker}}, \ and\ \bibinfo {author} {\bibfnamefont {Douglas}\ \bibnamefont
  {Stanford}},\ }\bibfield  {title} {\enquote {\bibinfo {title} {A bound on
  chaos},}\ }\href {\doibase 10.1007/JHEP08(2016)106} {\bibfield  {journal}
  {\bibinfo  {journal} {J. High Energ. Phys.}\ }\textbf {\bibinfo {volume}
  {2016}},\ \bibinfo {pages} {106} (\bibinfo {year} {2016})}\BibitemShut
  {NoStop}%
\bibitem [{\citenamefont {Roberts}\ \emph {et~al.}(2015)\citenamefont
  {Roberts}, \citenamefont {Stanford},\ and\ \citenamefont
  {Susskind}}]{roberts2015}%
  \BibitemOpen
  \bibfield  {author} {\bibinfo {author} {\bibfnamefont {Daniel~A.}\
  \bibnamefont {Roberts}}, \bibinfo {author} {\bibfnamefont {Douglas}\
  \bibnamefont {Stanford}}, \ and\ \bibinfo {author} {\bibfnamefont {Leonard}\
  \bibnamefont {Susskind}},\ }\bibfield  {title} {\enquote {\bibinfo {title}
  {Localized shocks},}\ }\href {\doibase 10.1007/JHEP03(2015)051} {\bibfield
  {journal} {\bibinfo  {journal} {J. High Energ. Phys.}\ }\textbf {\bibinfo
  {volume} {2015}},\ \bibinfo {pages} {51} (\bibinfo {year}
  {2015})}\BibitemShut {NoStop}%
\bibitem [{\citenamefont {Shenker}\ and\ \citenamefont
  {Stanford}(2014{\natexlab{a}})}]{shenker2014}%
  \BibitemOpen
  \bibfield  {author} {\bibinfo {author} {\bibfnamefont {Stephen~H.}\
  \bibnamefont {Shenker}}\ and\ \bibinfo {author} {\bibfnamefont {Douglas}\
  \bibnamefont {Stanford}},\ }\bibfield  {title} {\enquote {\bibinfo {title}
  {Black holes and the butterfly effect},}\ }\href {\doibase
  10.1007/JHEP03(2014)067} {\bibfield  {journal} {\bibinfo  {journal} {J. High
  Energ. Phys.}\ }\textbf {\bibinfo {volume} {2014}},\ \bibinfo {pages} {67}
  (\bibinfo {year} {2014}{\natexlab{a}})}\BibitemShut {NoStop}%
\bibitem [{\citenamefont {Shenker}\ and\ \citenamefont
  {Stanford}(2014{\natexlab{b}})}]{shenker2014a}%
  \BibitemOpen
  \bibfield  {author} {\bibinfo {author} {\bibfnamefont {Stephen~H.}\
  \bibnamefont {Shenker}}\ and\ \bibinfo {author} {\bibfnamefont {Douglas}\
  \bibnamefont {Stanford}},\ }\bibfield  {title} {\enquote {\bibinfo {title}
  {Multiple shocks},}\ }\href {\doibase 10.1007/JHEP12(2014)046} {\bibfield
  {journal} {\bibinfo  {journal} {J. High Energ. Phys.}\ }\textbf {\bibinfo
  {volume} {2014}},\ \bibinfo {pages} {46} (\bibinfo {year}
  {2014}{\natexlab{b}})}\BibitemShut {NoStop}%
\bibitem [{\citenamefont {Shenker}\ and\ \citenamefont
  {Stanford}(2015)}]{shenker2015}%
  \BibitemOpen
  \bibfield  {author} {\bibinfo {author} {\bibfnamefont {Stephen~H.}\
  \bibnamefont {Shenker}}\ and\ \bibinfo {author} {\bibfnamefont {Douglas}\
  \bibnamefont {Stanford}},\ }\bibfield  {title} {\enquote {\bibinfo {title}
  {Stringy effects in scrambling},}\ }\href@noop {} {\bibfield  {journal}
  {\bibinfo  {journal} {arXiv:1412.6087 [hep-th]}\ } (\bibinfo {year}
  {2015})},\ \Eprint {http://arxiv.org/abs/1412.6087} {arXiv:1412.6087
  [hep-th]} \BibitemShut {NoStop}%
\bibitem [{\citenamefont {Berry}(1977)}]{Berry_1977}%
  \BibitemOpen
  \bibfield  {author} {\bibinfo {author} {\bibfnamefont {M~V}\ \bibnamefont
  {Berry}},\ }\bibfield  {title} {\enquote {\bibinfo {title} {Regular and
  irregular semiclassical wavefunctions},}\ }\href {\doibase
  10.1088/0305-4470/10/12/016} {\bibfield  {journal} {\bibinfo  {journal}
  {Journal of Physics A: Mathematical and General}\ }\textbf {\bibinfo {volume}
  {10}},\ \bibinfo {pages} {2083--2091} (\bibinfo {year} {1977})}\BibitemShut
  {NoStop}%
\bibitem [{\citenamefont {Muessel}\ \emph {et~al.}(2015)\citenamefont
  {Muessel}, \citenamefont {Strobel}, \citenamefont {Linnemann}, \citenamefont
  {Zibold}, \citenamefont {{Juli{\'a}-D{\'i}az}},\ and\ \citenamefont
  {Oberthaler}}]{muessel2015}%
  \BibitemOpen
  \bibfield  {author} {\bibinfo {author} {\bibfnamefont {W.}~\bibnamefont
  {Muessel}}, \bibinfo {author} {\bibfnamefont {H.}~\bibnamefont {Strobel}},
  \bibinfo {author} {\bibfnamefont {D.}~\bibnamefont {Linnemann}}, \bibinfo
  {author} {\bibfnamefont {T.}~\bibnamefont {Zibold}}, \bibinfo {author}
  {\bibfnamefont {B.}~\bibnamefont {{Juli{\'a}-D{\'i}az}}}, \ and\ \bibinfo
  {author} {\bibfnamefont {M.~K.}\ \bibnamefont {Oberthaler}},\ }\bibfield
  {title} {\enquote {\bibinfo {title} {Twist-and-turn spin squeezing in
  {{Bose}}-{{Einstein}} condensates},}\ }\href {\doibase
  10.1103/PhysRevA.92.023603} {\bibfield  {journal} {\bibinfo  {journal} {Phys.
  Rev. A}\ }\textbf {\bibinfo {volume} {92}},\ \bibinfo {pages} {023603}
  (\bibinfo {year} {2015})}\BibitemShut {NoStop}%
\bibitem [{\citenamefont {Lerose}\ \emph {et~al.}(2018)\citenamefont {Lerose},
  \citenamefont {Marino}, \citenamefont {{\v Z}unkovi{\v c}}, \citenamefont
  {Gambassi},\ and\ \citenamefont {Silva}}]{lerose2018}%
  \BibitemOpen
  \bibfield  {author} {\bibinfo {author} {\bibfnamefont {Alessio}\ \bibnamefont
  {Lerose}}, \bibinfo {author} {\bibfnamefont {Jamir}\ \bibnamefont {Marino}},
  \bibinfo {author} {\bibfnamefont {Bojan}\ \bibnamefont {{\v Z}unkovi{\v c}}},
  \bibinfo {author} {\bibfnamefont {Andrea}\ \bibnamefont {Gambassi}}, \ and\
  \bibinfo {author} {\bibfnamefont {Alessandro}\ \bibnamefont {Silva}},\
  }\bibfield  {title} {\enquote {\bibinfo {title} {Chaotic {{Dynamical
  Ferromagnetic Phase Induced}} by {{Nonequilibrium Quantum Fluctuations}}},}\
  }\href {\doibase 10.1103/PhysRevLett.120.130603} {\bibfield  {journal}
  {\bibinfo  {journal} {Phys. Rev. Lett.}\ }\textbf {\bibinfo {volume} {120}},\
  \bibinfo {pages} {130603} (\bibinfo {year} {2018})}\BibitemShut {NoStop}%
\end{thebibliography}%

\end{document}